\documentclass[aps,pra,twocolumn]{revtex4-1} 

\usepackage{verbatim}
\usepackage{dsfont} 
\usepackage{amsmath}
\usepackage{amsthm} 
\usepackage{enumerate}
\usepackage{graphicx}
\usepackage{mathtools}
\usepackage{bm} 
\usepackage{amssymb} 
\usepackage{blindtext}
\usepackage{tcolorbox}
\usepackage{appendix} 
\usepackage{hyperref} 
\usepackage[capitalize,nameinlink]{cleveref}
\hypersetup{
    colorlinks=true,
    linkcolor=green,
    citecolor=magenta,
    unicode=true,  
}
\usepackage{textpos}

\newtheorem{result}{Result}
\newtheorem{definition}{Definition}
\newtheorem{lemma}{Lemma}

\newtheorem{remark}{Remark}

\newtheorem{corollary}{Corollary}

\DeclareMathOperator{\tr}{Tr}

\usepackage{calc} 
\usepackage{accents}

\newcommand{\indep}{\perp \!\!\! \perp}

\DeclareMathOperator{\sgn}{sgn}
\usepackage[normalem]{ulem}

\usepackage[framemethod=tikz,linewidth=0.5mm,linecolor=black,backgroundcolor=gray!5,innertopmargin=2mm,innerbottommargin=2mm]{mdframed}

\begin{document}

\title{Fundamental connections between utility theories of wealth and information theory}

\author{Andr\'es F. Ducuara$^{1,2}$} 
\email[]{andres.ducuara@yukawa.kyoto-u.ac.jp}

\author{Paul Skrzypczyk$^{3,4}$}
\email[]{paul.skrzypczyk@bristol.ac.uk}

\affiliation{$^{1}$Yukawa Institute for Theoretical Physics, Kyoto University, Kitashirakawa Oiwakecho, Sakyo-ku, Kyoto 606-8502, Japan
\looseness=-1}

\affiliation{$^{2}$Center for Gravitational Physics and Quantum Information, Yukawa Institute for Theoretical Physics, Kyoto University
\looseness=-1} 

\affiliation{$^{3}$H.H. Wills Physics Laboratory, University of Bristol, Tyndall Avenue, Bristol, BS8 1TL, United Kingdom 
\looseness=-1}

\affiliation{$^{4}$CIFAR Azrieli Global Scholars program, CIFAR, Toronto, Canada\looseness=-1}

\date{\today}

\begin{abstract}
    We establish fundamental connections between utility theories of wealth from the economic sciences and information-theoretic quantities. In particular, we introduce operational tasks based on betting where \emph{both} gambler and bookmaker have access to side information, or betting tasks with \emph{double} side information for short. In order to characterise these operational tasks we introduce new conditional R\'enyi divergences, and explore some of their properties. Furthermore, we introduce an utility theory of \emph{wealth ratios}, and operationally interpret there the two-parameter $(q,r)$ generalised mutual information measure recently introduced by V. M. Ili\'c and I. V. Djordjevi\'c \cite{ID1}; it quantifies the advantage provided by \emph{side information} in betting tasks for utility theories of wealth ratios. Moreover, we show that the Ili\'c-Djordjevi\'c conditional entropy \cite{ID1} satisfies a type of generalised chain rule, which generalises that of Arimoto-R\'enyi. Finally, we address the implications of these results on the quantum resource theories of informative measurements and non-constant channels. Altogether, these results further help strengthening the bridge between the theory of expected utility from the economic sciences and Shannon's theory of information.
\end{abstract} 
\maketitle

\begin{textblock*}{3cm}(17cm,-10.5cm)
  \footnotesize YITP-23-68
\end{textblock*}
\vspace{-1cm}
\section{Introduction}

Information theory (IT) is a theoretical framework which deals with the manipulation, storage, and processing of information, and which was formalised by Shannon in 1948 \cite{shannon1}, with important earlier contributions made by Nyquist and Hartley during the 1920's \cite{nyquist, hartley}. One of the main objects of study here is that of \emph{entropic quantities}, which come in the form of: unconditional and conditional entropies, mutual information measures, capacities, divergences, conditional divergences, amongst others \cite{CT}. In addition to their role within IT itself, these quantities have been found to play fundamental roles in the development of scientific disciplines such as: thermodynamics, statistical mechanics, black hole thermodynamics, quantum gravity, quantum information theory,  biological and social sciences, amongst others \cite{CT}. It therefore becomes of foundational and practical importance to keep exploring the mathematical properties as well as the physical and operational significance of such information-theoretic quantities.

Expected utility theory (EUT) on the other hand, is a theoretical framework first formalised by von Neumann and Morgenstern within the theory of games and economic behaviour in 1944 \cite{risk_vNM}. In broad terms, EUT deals with the behaviour of rational agents when faced with decision problems. One of the main objects of study within EUT is that of the \emph{utility function} (of a rational agent), a real-valued function representing the agent's level of satisfaction when acquiring an amount of wealth, goods, or services. Utility functions are particularly useful at describing behavioural tendencies of rational agents such as the agent's aversion to risk. Another important object of study, and related to the utility function, is that of the \emph{certainty equivalent}. The certainty equivalent (for a given lottery) represents the certain amount of wealth, goods, or services, which the rational agent is willing to accept, so in order to walk away from participating in the given lottery. In other words, it is the \emph{certain} amount of wealth that is \emph{equivalent} (from the agent's point of view) to the lottery in question. The characterisation of agents' behavioural tendencies has arguably been the bread and butter, so to speak, of the economic sciences since their formal introduction during the 1950's and throughout the second half of the previous century, with many of these efforts being recognised with the Sveriges Riksbank Prize in Economic Sciences in Memory of Alfred Nobel. Whilst initially a concept of interest mostly to the economic sciences, the behavioural tendencies of rational agents turned out to be a ubiquitous concept that has emerged and found usefulness in other scientific disciplines such as behavioural ecology, neuroscience, and information theory, with this latter case being the main focus of this manuscript. 

One of the earliest examples of the interplay between expected utility theory and information theory, to the best of our knowledge, was addressed in the work of Kelly in 1956 \cite{kelly}. In this work, Kelly formalised and characterised the operational tasks of betting on horse races, or horse betting for short and, in particular, addressed a notion of wealth which was found to be related to Shannon's information-theoretic quantities such as the entropy and the mutual information \cite{kelly, CT}. Whilst the concept of risk-aversion was not explicitly addressed as an ingredient in Kelly's original proposal, it is now understood that the scenario introduced by Kelly corresponds to a risk-averse gambler with constant relative risk aversion (CRRA) given by $R=1$ (relative risk aversion being one way to quantify the agent's aversion to risk). Fast forward to the twenty-first century, in 2020 Bleuler, Lapidoth, and Pfister (BLP) took an important step in \cite{BLP1}, by considering horse betting more generally for gamblers with risk-aversion values spanning $R\in \mathds{R}_{\geq 0}\cup\{+\infty\}$. In addition to this, BLP introduced a new conditional R\'enyi divergence and a new mutual information measure, which helped characterising horse betting scenarios where the gambler has access to side information \cite{BLP1}. Later on, another step was taken in 2022 in \cite{DS2022}, where the the authors introduced a framework for: negative wealth, risk-aversion values spanning the whole extended line of real numbers $R\in \mathds{R}\cup\{+\infty, - \infty\}$, as well as explicitly identifying the role of the \emph{certainty equivalent} in horse betting. The authors in \cite{DS2022} also showed that Arimoto's mutual information measure quantifies the usefulness that side information provides for these tasks, and extended horse betting to the quantum domain, where families of additional \emph{betting tasks} naturally emerge: quantum state betting, quantum channel/subchannel betting, amongst other variants. The current manuscript aims to be a spiritual successor of this line of research.

In this work we further explore and expand the scope of the connections between expected utility theory and information theory by showing that the above-mentioned cases are not isolated examples, but that these two theories are further intimately connected at a fundamental level. We do this by uncovering additional scenarios where both expected utility theory and information theory jointly describe decision problems and betting tasks for utility theories of wealth. Our main findings are the following.

First, within the utility theory of \emph{wealth}, we address betting tasks where \emph{both} gambler and bookmaker have access to side information, or betting tasks with \emph{double} side information for short. The exploration of these operational tasks leads us to the identification of new R\'enyi conditional divergences which precisely characterise the advantage provided by side information in such betting tasks. In particular, we find that side information being available to the bookmaker can actually help the gambler, but that this advantage cannot be larger than when the gambler has direct access to such side information. Second, we further generalise these results to the so-called \emph{original prospect theory} (OPT), a generalised version of EUT, in which decision-making agents are allowed to behave irrationally, albeit in a systematic manner \cite{PT1}. In this regard, we  prove that a specific  R\'enyi conditional entropy characterises such deviation from rationality. Third, we introduce an utility theory of \emph{wealth ratios} and prove that the recently introduced Ili\'c-Djordjevi\'c measure of mutual information \cite{ID1} quantifies the advantage provided by \emph{side information} in betting tasks for such an utility theory. Moreover, we prove that the Ili\'c-Djordjevi\'c conditional entropy \cite{ID1} satisfies a type of generalised chain rule, which generalises that of Arimoto-R\'enyi. Fourth and finally, we address the implications of these results, about the utility theory of wealth ratios, on quantum state betting games (QSB) within the quantum resource theory (QRT) of informative measurements, as well as noisy QSB games within the QRT of non-constant channels.

This work is organised as follows. In Sec.~\ref{s:background} we start with preliminaries of the background theories. In Sec.~\ref{ss:IT} we address some information-theoretic quantities and introduce new conditional R\'enyi divergences, whilst in Sec.~\ref{ss:EUT} we describe expected utility theory, the concept of risk-aversion, the certainty equivalent, and prospect theory. In Sec.~\ref{s:R} we start our results sections. In Sec.~\ref{ss:BT} we address operational tasks based on betting, risk-aversion, and double side information. In Sec.~\ref{ss:BT2} we characterise betting tasks with double side information. In Sec.~\ref{ss:PT} we consider a generalisation to original prospect theory. In Sec.~\ref{ss:WR} we introduce an utility theory framework for wealth ratios, and in Sec.~\ref{ss:WR2} a characterisation of betting tasks in such a scenario. In Sec.~\ref{ss:R45} we address the implications of these results on the QRTs of informative measurements and non-constant channels. We finish in Sec.~\ref{s:conclusions} with conclusions.

\section{Preliminaries}
\label{s:background}

In this section we address some generalised entropic measures, the theory of expected utility, the concept of risk-aversion, risk aversion measures, the certainty equivalent, and prospect theory.

\subsection{Information-theoretic quantities}
\label{ss:IT}
 
One important generalisation of the entropic quantities \`a la Shannon \cite{shannon1} is the extension due to R\'enyi \cite{renyi}. In this direction, the Shannon entropy and the Kullback-Leibler (KL) divergence (or KL relative entropy) are generalised as the R\'enyi entropy and the R\'enyi divergence, respectively \cite{renyi, RD}. Interestingly however, there are instead various R\'enyi alternatives for the conditional entropy \cite{review_RCE}, as well as for the mutual information such as the proposals of: Arimoto \cite{arimoto}, Sibson \cite{sibson}, Csisz\'ar \cite{csiszar}, Lapidoth-Pfister \cite{LP}, Tomamichel-Hayashi \cite{TH}, amongst others \cite{ID1}. Whilst all of these mutual information measures have found usefulness within information theory \cite{ID1}, Arimoto's measures stand out as strong candidates due to them satisfying a considerable number of desirable properties \cite{review_RCE}.

The previous addressed extension \`a la R\'enyi is however not the only way to generalise Shannon's entropy. The Tsallis-Havrda-Charv\'at entropy, first introduced in the context of information theory by Havrda and Charv\'at \cite{Havrda_Charvat}, and later on independently introduced in the realm of statistical physics by Tsallis \cite{Tsallis1}, is a non-additive generalisation of Shannon's.  More generally, the two parameter $(q,r)$ entropy introduced by Sharma and Mittal  \cite{SM1} recovers both R\'enyi and Tsallis in the limits $(q,1)$ and $(q,q)$, respectively. Similarly to the R\'enyi case however, whilst there is a generally accepted unique way of introducing a unconditional entropy \`a la Tsallis as well as \`a la Sharma-Mittal, here again one important challenge is the generalisation of additional information-theoretic quantities. In this regard, an important step was recently taken in the work of Ili\'c and Djordjevi\'c \cite{ID1}, where various entropic measures \`a la Sharma-Mittal were introduced: a conditional entropy, a mutual information measure, and a channel capacity. In particular, the Ili\'c-Djordjevi\'c measures reduce to the Arimoto-R\'enyi conditional entropy, Arimoto's mutual information, and Arimoto-R\'enyi capacity, in the limit $(q,1)$. Furthermore, these measures happen to satisfy a considerable number of desirable properties \cite{ID1} and consequently, present themselves as strong generalisations of Shannon's entropic quantities in the direction of Sharma-Mittal. 

We now introduce the unconditional entropy of Sharma-Mittal, the conditional entropy and mutual information of Ili\'c and Djordjevi\'c, the R\'enyi divergence, a conditional R\'enyi divergences, and a R\'enyi conditional entropy. We use the following notation: We consider random variables ($X, G,...$) on a finite alphabet $\mathcal{X}$, and the probability mass function (PMF) of $X$ represented as $p_X$ satisfying: $p_X(x)\geq 0$, $\forall x \in \mathcal{X}$, and $\sum_{x \in \mathcal{X}} p_X(x) = 1$. We omit the alphabet when summing, and write $p_X(x)$ as $p(x)$ when evaluating. We denote the support of $p_X$ by ${\rm supp} (p_X) \coloneqq \{x\,|\,p(x)>0\}$, the cardinality of the support by $|{\rm supp}(p_X)|$, the set of non-negative real numbers by $\mathds{R}_{\geq 0}$, and the extended line of real numbers by $ \mathds{ \overline R}\coloneqq \mathds{R} \cup \{\infty,-\infty\}$. Joint and conditional PMFs are denoted by $p_{XG}$, $p_{G|X}$ respectively, and we also consider the auxiliary ``sign" function:
\begin{align}
    \sgn(\alpha)
    \coloneqq
    \begin{cases}
        +1, & \alpha \geq 0,
        \\
        -1,& \alpha < 0.
    \end{cases}
    \label{eq:sgn}
\end{align}
We start by considering the Sharma-Mittal entropy.

\begin{definition} (Sharma-Mittal entropy \cite{SM1})
	The Sharma-Mittal entropy of order $(q,r) \in \mathds{\overline R} \times \mathds{R}$ of a PMF $p_X$ is denoted as $H^{\rm SM}_{q,r} (X)$. The orders $q,r \in (-\infty,0) \cup (0,1) \cup (1,\infty)$ are defined by:
	\begin{align}
		H^{\rm SM}_{q,r}
		(X)
		& \coloneqq
		\frac{
            \sgn(q)
            }{1-r}
		\left[
		\left(
		\sum_x
		p(x)^q
		\right)^{\frac{1-r}{1-q}}
		-1
		\right]
		.
		\label{eq:SME}
	\end{align}
	The remaining orders are defined by continuous extension of \eqref{eq:SME}. In particular, the Sharma-Mittal entropy recovers Tsallis and R\'enyi entropies in the limits $(q, r\rightarrow q)$ and $(q, r\rightarrow 1)$, respectively.
\end{definition}
The Sharma-Mittal entropy has been explored in the context of statistical physics as well as information theory \cite{masi1, masi2, sm2, sm3}. In addition to recovering both R\'enyi $(q, r\rightarrow 1)$ and Tsallis $(q, r\rightarrow q)$, the limit $(q\rightarrow 1, r)$ (called \emph{Gaussian entropy}) has also been explored independently \cite{g1, g2, g3} and, similarly, the so-called Landsberg-Vedral entropy $(q,2-q)$ \cite{lv1, lv2, lv3}. We now move on to the information theoretic quantities, based upon the Sharma-Mittal entropy, recently introduced by Ili\'c and Djordjevi\'c \cite{ID1}.

\begin{definition} (Ili\'c-Djordjevi\'c information theoretic quantities \cite{ID1}) The Ili\'c-Djordjevi\'c conditional entropy and mutual information measures of order $(q,r)\in \mathds{\overline R} \times \mathds{R}$ of a joint PMF $p_{XG}$ are denoted by $H_{q,r}^{\rm ID} (X|G)$ and $I^{\rm ID}_{q,r}(X;G)$ respectively. The orders $q,r \in (-\infty,0) \cup (0,1) \cup (1,\infty)$ are defined by:
{\small
\begin{align}
    H^{\rm ID}_{q,r}(X|G)
    &\coloneqq
    \frac{
        \sgn(q)
    }{
        1-r
    }
    \left[
    \left(
    \sum_g
    \left(
    \sum_x p(x,g)^q
    \right)^\frac{1}{q}
    \right)^{\hspace{-0.2cm}\frac{q(1-r)}{(1-q)}}
    \hspace{-0.4cm}
    -
    1
    \right]
    ,
    \label{eq:IDCE}\\
    I^{\rm ID}_{q,r}(X;G)
    &\coloneqq
    H^{\rm SM}_{q,r}(X)
    \ominus_r
    H^{\rm ID}
    _{q,r}(X|G)
    ,
    \label{eq:IDMI}
\end{align}}
where the pseudo-subtraction operation $\ominus_r$ is given by $x \ominus_r y\coloneqq \frac{x-y}{1+(1-r)y}$. These quantities recover the Arimoto-R\'enyi conditional entropy and the Arimoto-R\'enyi mutual information respectively, in the limit $(q, r\rightarrow 1)$. In particular, the case $(q,r)=(1,1)$ recovers the standard mutual information $I^{\rm ID}_{1,1}(X;G) =  H(X) - H(X|G)$ \cite{CT}. The remaining orders are defined by continuous extension of \eqref{eq:IDCE}.
\end{definition}

Given this two-parameter mutual information measure, we can naturally define mutual information measures of Arimoto-Tsallis $(q, r\rightarrow q)$, Arimoto-Gauss $(1,r)$, in the corresponding limits, following the same limit cases as for the unconditional measure. The Ili\'c-Djordjevi\'c mutual information satisfies various desirable properties which makes it a strong candidate amongst several other proposals \cite{ID1}. We now introduce the R\'enyi divergence, a conditional R\'enyi divergence, and a R\'enyi conditional entropy. 

\begin{definition} (Some R\'enyi information-theoretic quantities)
	The R\'enyi divergence (R-divergence) of order $\alpha \in \mathds{\overline R}$ of PMFs $p_X$ and $q_X$ is denoted as $D_\alpha(p_X||q_X)$ \cite{renyi, RD}. The Bleuler-Lapidoth-Pfister conditional-R\'enyi divergence (BLP-CR-divergence) of order $\alpha \in \mathds{\overline R}$ of conditional PMFs $p_{X|G}$, $q_{X|G}$, and PMF $p_X$ is denoted as $D^{\rm BLP}_\alpha
	(
	p_{G|X}
	||
	q_{G|X}
	|
	p_X
	)$ \cite{BLP1, thesis_CP}. The number two R\'enyi conditional entropy of order $\alpha \in \mathds{\overline R}$ of a joint PMF $p_{XG}$ is denoted as $H_{\alpha}^{2}(X|G)$, following the notation by Fehr and Berens \cite{review_RCE}. The orders $\alpha \in (-\infty,0)\cup(0,1)\cup(1,\infty)$ of these quantities are defined as:
	\begin{align}
    	&D_\alpha(p_X||q_X)
    	\coloneqq
    	\frac{
    	\sgn(\alpha)
    	}{
    	\alpha-1
    	}
    	\ln
    	\left[
    	\sum_x
    	p(x)^\alpha
    	q(x)^{1-\alpha}
    	\right]
    	,
    	\label{eq:RD}
     \\
     &D^{\rm BLP}_\alpha
	(
	p_{X|G}
	||
	q_{X|G}
	|
	p_G
	)
	\label{eq:BLPCRD}
	\\
	&\coloneqq
	\frac{|\alpha|}{\alpha-1}
	\ln
	\left[
    	\sum_g
    	p(g)
    	\left(
        	\sum_x
        	p(x|g)^\alpha
        	q(x|g)^{1-\alpha}
    	\right)^{\frac{1}{\alpha}}	
	\right]
	,
	\nonumber
 \\
 &H_{\alpha}^2
	(X|G)
	\coloneqq
	H_\alpha^{\rm R}
	(X,G)
	-
	H_\alpha^{\rm R}
	(G)
        ,
	\label{eq:2RCE}
	\end{align}
with $H_\alpha^{\rm R}(\cdot)$ the R\'enyi entropy. The orders $\alpha \in\{1,0,\infty,-\infty\}$ are defined define by their respective continuous extensions.
\end{definition}

Further details about the relationship between the R\'enyi divergence, conditional R\'enyi divergences, and R\'enyi conditional entropies in \cref{a:relationships}. We now introduce two new conditional R\'enyi divergences, address some of their properties, as well as the way they relate to other information-theoretic quantities.

\begin{definition} (First new conditional R\'enyi divergence) 
	The first new conditional R\'enyi divergence (n1-CR-divergence) of order $\alpha \in \mathds{\overline R}$ of PMFs $p_{X|Y}$, $q_{X|Y}$, $p_Y$ is denoted as 
	$
	D^{\rm n1}_\alpha
	(
	p_{X|Y}
	||
	q_{X|Y}
	|
	p_{Y}
	)
	$. The orders $\alpha\in(-\infty,0)\cup(0,1)\cup(1,\infty)$ are defined as:
	\begin{align}
	&D^{\rm n1}_\alpha
	(
	p_{X|Y}
	||
	q_{X|Y}
	|
	p_{Y}
	)
	\label{eq:n1CRD}
	\\
	&=
	\frac{\sgn(\alpha)}{\alpha-1}
	\ln
	\left[
        	\sum_x
        	    \left(
        	    \sum_y
        	        p(y)\,
                	p(x|y)\,
                	q(x|y)^\frac{1-\alpha}{\alpha}
            	\right)^\alpha
	\right]
	.
	\nonumber
	\end{align}	
	The orders $\alpha \in\{1,0,\infty,-\infty\}$ are defined by continuous extension of \eqref{eq:n1CRD}.
\end{definition}
We can check that this quantity satisfies two natural properties which are expected from a CR-divergence: i) non-negativity, and ii) reduction to the R\'enyi divergence. The latter means that when $X \indep Y$ ($X$ independent of $Y$) we recover the R\'enyi divergence as: 
$
D^{\rm n1}_\alpha
(
p_{X|Y}
||
q_{X|Y}
|
p_{Y}
)
=
D_\alpha
(
p_{X}
||
q_{X}
)
$. Another desirable property to have is the data processing inequality.

\begin{lemma}
\label{r:r1}
    (Data processing inequality for the n1-CR-divergence)
    Consider conditional PMFs $p_{X|Y}$, $q_{X|Y}$, PMF $p_Y$, and orders $\alpha \in (0,1)$, then:
    \begin{align}
        D_\alpha
    	(
    	p_{X}
    	||
    	q_{X}
    	)
    	\leq
        D^{\rm n1}_\alpha
    	(
    	p_{X|Y}
    	||
    	q_{X|Y}
    	|
    	p_{Y}
    	)
    	,
    \end{align}
    with
    the PMFs $p_X$, $q_X$ given by 
    $
     p(x)
    \coloneqq
    \sum_y
    p(x|y) 
    p(y)
    $,
    $
    q(x)
    \coloneqq
    \sum_y
    q(x|y) 
    p(y)
    $.
\end{lemma}

The proof of this lemma is in \cref{a:r1}. It will also prove useful to consider the following inequality between the n1-CR-divergence and the BLP-CR-divergence.

\begin{lemma}\label{r:r2}
    (Inequality between the n1-CR divergence and the BLP-CR divergence) Consider conditional PMFs $p_{X|Y}$, $q_{X|Y}$, PMF $p_Y$, and orders $\alpha\in(-\infty,0)\cup(0,1)\cup(1,\infty)$, we have:
    \begin{align}
        D^{\rm n1}_\alpha
    	(
    	p_{X|Y}
    	||
    	q_{X|Y}
    	|
    	p_{Y}
    	)
    	\leq
        D^{\rm BLP}_\alpha
    	(
    	p_{X|Y}
    	||
    	q_{X|Y}
    	|
    	p_{Y}
    	)
    	,
    \end{align}
    with the n1-CR divergence \eqref{eq:n1CRD} and the BLP-CR divergence \eqref{eq:BLPCRD}.
\end{lemma}
The proof of this lemma is in \cref{a:r2}. We now introduce a second information-theoretic quantity which is  more general than the described above.

\begin{definition} (Second new conditional R\'enyi divergence) 
    The second new conditional R\'enyi divergence (n2-CR-divergence) of order $\alpha \in \mathds{\overline R}$ of PMFs $p_{X|GY}$, $q_{X|GY}$, $p_G$, and $p_{Y|G}$ is denoted as $D^{\rm n2}_\alpha
	(
	p_{X|GY}
	||
	q_{X|GY}
	|
	p_{G}
	,
	p_{Y|G}
	)$. The orders $\alpha\in(-\infty,0)\cup(0,1)\cup(1,\infty)$ are defined as:
	{\small\begin{align}
	&
	D^{\rm n2}_\alpha
	(
	p_{X|GY}
	||
	q_{X|GY}
	|
	p_{G}
	,
	p_{Y|G}
	)
	\coloneqq
	\frac{|\alpha|}{\alpha-1}
	\label{eq:n2CRD}
	\\
	&
	\ln
	\left[
    	\sum_g
    	p(g)
    	\left(
        	\sum_x
        	    \left(
        	    \sum_y
        	        p(y|g)
        	        \,
                	p(x|gy)
                    \,	q(x|gy)^\frac{1-\alpha}{\alpha}
            	\right)^\alpha
    	\right)^{\frac{1}{\alpha}}	
	\right]
	.
	\nonumber
	\end{align}}
	The orders $\alpha \in\{1,0,\infty,-\infty\}$ are defined by continuous extension of \eqref{eq:n2CRD}.
\end{definition}
We can check that this quantity is related the previously defined n1-CR-divergence, as well as the BLP-CR-divergence and the R\'enyi divergence, as follows.

\begin{remark}
    (Relationship between the n2-CR divergence, n1-CR divergence, and the R\'enyi divergence) Considering the dependence/independence between random variables $X$, $G$, and $Y$, we can have the following equalities:
    {\small\begin{align}
        D^{\rm n2}_\alpha
    	(
    	p_{X|GY}
    	||
    	q_{X|GY}
    	|
    	p_{G}
    	,
    	p_{Y|G}
    	)
    	&=
    	D^{\rm n1}_\alpha
    	(
    	p_{X|Y}
    	||
    	q_{X|Y}
    	|
    	p_{Y}
    	)
    	,
    	\\
    	D^{\rm n2}_\alpha
    	(
    	p_{X|GY}
    	||
    	q_{X|GY}
    	|
    	p_{G}
    	,
    	p_{Y|G}
    	)
    	&=
    	D^{\rm BLP}_\alpha
    	(
    	p_{X|G}
    	||
    	q_{X|G}
    	|
    	p_{G}
    	)
    	,
    	\\
    	D^{\rm n1}_\alpha
    	(
    	p_{X|Y}
    	||
    	q_{X|Y}
    	|
    	p_{Y}
    	)
    	&=
    	D_\alpha
    	(
    	p_{X}
    	||
    	q_{X}
    	)
    	,
    	\\
    	D^{\rm BLP}_\alpha
    	(
    	p_{X|G}
    	||
    	q_{X|G}
    	|
    	p_{G}
    	)
    	&=
    	D_\alpha
    	(
    	p_{X}
    	||
    	q_{X}
    	)
    	.
    \end{align}}
    The first equality holds when  $(X,Y) \indep G$, the second and third equalities when $X \indep Y$, and the fourth equality when  $X \indep G$.
\end{remark}

\begin{figure}[h!]
    \centering
    \includegraphics[scale=0.58]{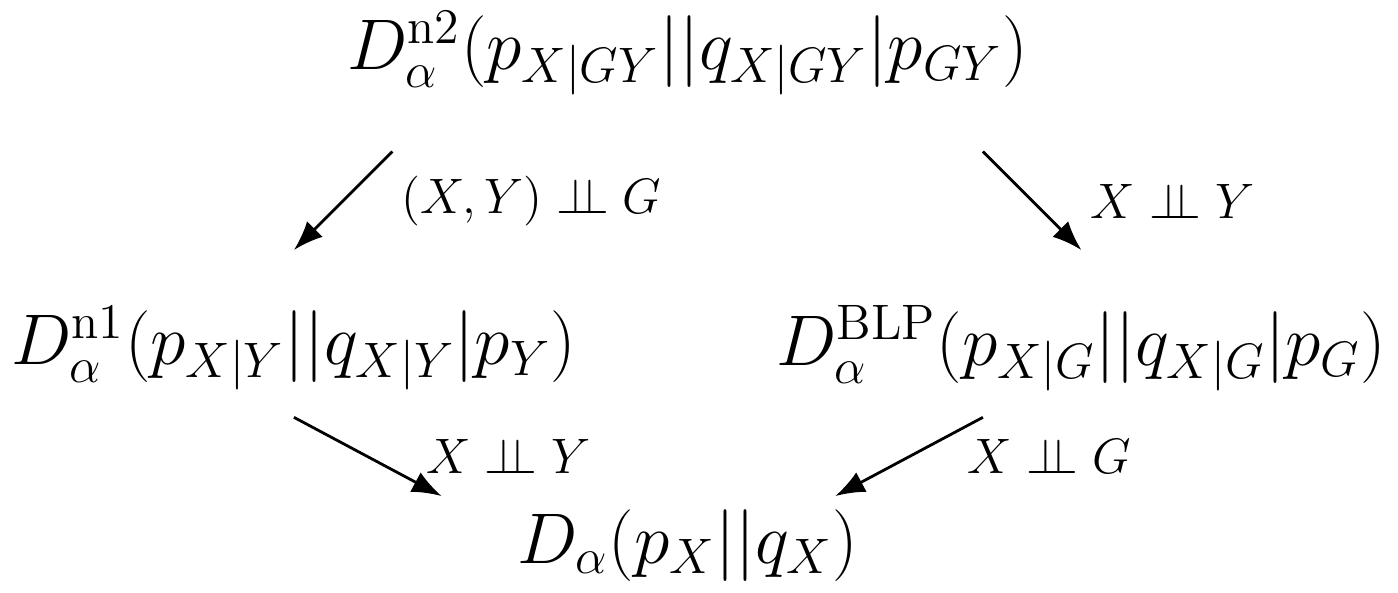}
    \vspace{-0.3cm}
    \caption{Relation between the new conditional R\'enyi divergences, the BLP-CR divergence, and the R\'enyi divergence. $A \overset{C}{\rightarrow} B$ means that ``quantity $A$ becomes equal to quantity $B$ when condition $C$ holds". The notation $X \indep Y$ means that $X$ is independent of $Y$.}
    \label{fig:CR_divergences}
\end{figure}

The previous remarks and results are summarised in \autoref{fig:CR_divergences}. We now move on to the description of expected utility theory and prospect theory from the economic sciences.  

\subsection{Expected utility theory, risk-aversion, the certainty equivalent, and prospect theory}
\label{ss:EUT}

The theory of expected utility, first formalised in the work of von Neumann and Morgenstern \cite{risk_vNM}, deals with rational agents and their prospects of earning, losing, hoarding, and interchanging goods like wealth, services, and similar assets. A central object of study here is the concept of \emph{utility function}, a real-valued function $u:A\rightarrow \mathds{R}$, with $A$ a set of ``alternatives" endowed with a binary relation $\prec$. The general idea is for the utility function to represent the \emph{level of satisfaction} of a rational agent  ($u(a)\in \mathds{R}$) with the prospect of acquiring a specific alternative ($a\in A$). The binary relation $a \prec b$ can be regarded as ``alternative $b$ represents more wealth or more services than alternative $a$". Now, it is then natural for rational agents to be more pleased with $b$ than with $a$, and so this should be reflected in the agent's utility function as $u(a)< u(b)$, which therefore imposes the utility function to be a monotone for the binary relation. This is an appealing way of formalising utility, since it is natural for a rational agent to regard more goods or wealth, as something generally better than less goods or wealth.

In this work we deal with the set of alternatives being either \emph{wealth}, or \emph{wealth ratios} (to be introduced later on in the document), and so it is enough to consider an interval in the real numbers $A\subseteq \mathds{R}$, with the binary relation being the standard relation $<$. In this section we then address the quintessential example of an utility theory of \emph{wealth}. The utility function being monotonic gets translated into being an strictly increasing function and, additionally, we ask for it to be twice-differentiable, for mathematical convenience, and because it is natural to consider that small changes in wealth imply small changes in the agent's satisfaction. We note here that the utility function does not necessarily need to be positive or negative, since the idea is for it to compare alternatives only. We also note that whilst one could exclusively work with positive wealth, one can also construct scenarios where wealth is negative, and so the utility function would  then naturally represent the level of (dis)satisfaction of the rational agent when having to pay or give back the amount of wealth in question \cite{DS2022}.

Another important aspect of the theory of expected utility is the behaviour of a rational agent in the presence of uncertainty. Consider an random event distributed according to the PMF $p_X$ on a finite alphabet $\mathcal{X}$, and a distribution of wealth given by $w_X$, with $w(x) \in \mathds{R}_{\geq 0}$, $\forall x\in \mathcal{X}$.  The pair $(p_X, w_X)$ is often refereed to as a \emph{lottery}. Fix a lottery $(p_X, w_X)$ and consider a game where a rational agent, represented by a utility function $u$, is asked to predict outcome $x\in \mathcal{X}$, which occurs with probability $p(x)$, and for which is rewarded (after correctly guessing) an amount of wealth $w(x)\in \mathds{R}_{\geq 0}$. The \emph{expected utility} of the rational agent playing such a game is given by $\mathbb{E}_{p_X}(u(w_X)) = \sum_x u(w(x)) p(x)$. Using this expected utility, a central object of interest is that of the \emph{certainty equivalent (CE)}. The CE is defined as the amount of wealth $w^{\rm CE}$ which has an utility that exactly matches the expected utility of the lottery as:
\begin{align}
    u
    \left(
    w^{\rm CE}
    \right)
    =
    \mathbb{E}_{p_X}
    \left[
    u
    \left(
    w_X
    \right)
    \right]
    .
    \label{eq:CE1}
\end{align}
Qualitatively, the CE represents how attractive the lottery is to the rational agent. In other words, it represents the minimum \emph{certain} amount of wealth that the rational agent is willing to receive, so to get persuaded of \emph{not} playing the game. We remark here that the CE is a quantity that depends both on the lottery $(p_X, w_X)$ as well as on the agent's utility function $u$, and so we can explicitly write this as:
\begin{align}
    w^{\rm CE}_u (p_X, w_X)
    &=
    u^{-1}
         \left(
     \mathbb{E}_{p_X}
    (
    u(w_X)
    )
        \right)
    .
    \label{eq:wCE2}
\end{align}
The CE stands out as an important quantity because optimising it is equivalent to optimising the \emph{expected utility}, due to the utility function being strictly increasing and because $u(w^{\rm CE}) = \mathds{E}_{p_X}[u(w_X)]$. Taking the latter into account, and that the CE has units of wealth $[w]$ (\$, \pounds, \textyen, ...), it is usually better to consider the CE, instead of the expected utility, as the figure of merit for the setups involving rational agents placing bets that we will address later on. In addition to this, the CE helps establishing the characterisation of risk tendencies of rational agents \cite{DS2022}. Using the CE, the concept of risk-aversion in the context of utility theory emerges via the highly nontrivial realisation that the risk-attitude of a rational agent is related to the concavity (risk-averse), convexity (risk-seeking), or linearity (risk-neutral) of the agent's utility function  \cite{risk_bernoulli, risk_arrow, risk_pratt, risk_finetti} (a detailed derivation for both positive and negative wealth is in \cite{DS2022}). One common measure of risk-aversion is the so-called relative risk aversion (RRA) measure given by \cite{risk_arrow, risk_pratt, risk_finetti}:
\begin{align}
    RRA_u
    (
    w
    )
    \coloneqq
    -
    w
    \frac{
    u^{''}
    (w)
    }{
    u'(w)
    }
    .
    \label{e:RRA}
\end{align}
This quantifier is dimensionless, which is a characteristic not satisfied by all quantifiers of risk-aversion \cite{risk_arrow, risk_pratt, risk_finetti}. The RRA measure does not assign a global value for risk aversion, so we ask here for utility functions where the agent's RRA is \emph{constant}. We can solve \eqref{e:RRA} assuming $RRA(w) = R$, leading to the constant relative risk aversion (CRRA) function, or \emph{isoelastic} utility function, for both positive and negative wealth as:
\begin{align}
    u_R^{\rm I}
    (w)
    =
    \sgn(w)
    \ln_R(
    |w|
    )
    ,
    \label{eq:E}
\end{align}  
with the auxiliary ``sign" function \eqref{eq:sgn} and  the $R$-deformed natural logarithm as:
\begin{align}
    \ln_R
    (x)
    &\coloneqq
    \begin{cases}
    \ln x, 
    & 
    R=1,
    \\
    \frac{
    x^{1-R}
    -
    1
    }{
    1-R
    }
    ,
    & 
    R\neq 1,
\end{cases}
\end{align}
with the deformation parameter $R \in \mathds{R}$. The parameter $R$ varies from minus to plus infinity, describing all possible risk tendencies of the rational agent, for either positive or negative wealth \cite{DS2022}. The certainty equivalent of the isoelastic function, or \emph{isoelastic certainty equivalent} (ICE), the figure of merit in this work, is given by: 
\begin{align}
    w^{\rm ICE}_R
    (p_X, w_X)
    &\coloneqq
    (u_R^{\rm I})^{-1}
         \left(
     \mathbb{E}_{p_X}
    (
    u_R^{\rm I}
    (w_X)
    )
        \right)
    .
    \label{eq:CE2}
\end{align}

We now address a generalisation of expected utility theory which goes under the name of \emph{prospect theory} \cite{PT1}. The original version of this theory, here addressed as original prospect theory (OPT), was first introduced by Kahneman and Tversky in 1979 \cite{PT1}, as a generalisation of expected utility theory where, amongst other features, decision-making agents do not necessarily have to behave rationally in regards to the assessment of probabilities. More explicitly, this means that when dealing with decision problems, decision-making agents do not necessarily regard probabilities as $\{p(x)\}$, but instead treat probabilities in a distorted manner as $\{ \pi(p(x)) \}$, with $\pi:[0,1]\rightarrow [0,1]$ a \emph{probability weighting function} \cite{PT_review_2013, PT_review_2010, PT_review_2004, PT_review_2000}. This generalisation better matches experimental data and has more explanatory power than EUT. Later on, OPT was further generalised into cumulative prospect theory (CPT) in 1992 \cite{PT2}. In general, the descriptive power of prospect theory has had profound implications in various research fields, most notably within the discipline of behavioural economics. These efforts were acknowledged with the Sveriges Riksbank Prize in Economic Sciences in Memory of Alfred Nobel in 2002 to Daniel Kahneman and Vernon L. Smith (Tversky passed away in 1996 and Nobel prizes are not awarded posthumously). These two variants of prospect theory, OPT and CPT, are areas of active research in modern economic sciences \cite{GW2022, pan2019, mohammed2010}. We now proceed to describe OPT in more detail. 

In original prospect theory (OPT), the \emph{expected utility} of a lottery $(p_X, w_X)$, for a decision-making agent represented by the utility function $(u)$, is generalised to a \emph{value function}, where the decision-making agent is represented instead by a \emph{pair} of utility function and probability weighting function $(u, \pi)$. The value function in OPT, denoted here as $v_{u,\pi}^{\rm PT}$, is calculated in OPT according to the composition rule given by:
\begin{align}
    v^{\rm PT}_{u,\pi}
    (
    p_X
    ,
    w_X
    )
    \coloneqq
    \sum_x
    \pi
    (
    p(x)
    )
    \,
    u
    (w(x))
    .
\end{align}
This value function is meant to preserve the meaning behind that of the expected utility, in the sense that decision-making agents would now behave as \emph{value function maximisers} (as opposed to expected utility maximisers), meaning that agents would prefer the lottery which \emph{maximises} their value function. Taking into account that the value function is still a linear combination of the utilities $\{u(w(x))\}$ (so it still has units of ``utility of wealth"), we can still define a certainty equivalent of wealth now for prospect theory as:
\begin{align}
    w^{\rm CE-PT}_{u,\pi}
         (
         p_X
         ,
         w_X
         )
         \coloneqq
         u^{-1}
         \left(
     v^{\rm PT}_{u,\pi}
        (
        p_X
        ,
        w_X
        )
        \right)
        .
\end{align}
As in EUT, the CE for OPT represents the value (in units of wealth) that a decision-making agent (represented by the pair $(u,\pi)$) assigns to the lottery $(p_X,w_X)$. In particular, we can naturally recover both the expected utility and the standard CE by considering that the decision-making agent behaves completely rational with respect to probabilities $\pi(\cdot)={\rm id}(\cdot)$ as:
\begin{align}
    v^{\rm PT}_{u,\pi={\rm id}}
    (
    p_X
    ,
    w_X
    )
    &=
    \mathbb{E}_{p_X}
    (
    u(w_X)
    )
    ,
    \\
    w^{\rm CE-PT}_{u,\pi={\rm id}}
         (
         p_X
         ,
         w_X
         )
    &=
    u^{-1}
         \left(
     \mathbb{E}_{p_X}
    (
    u(w_X)
    )
        \right)
        .
\end{align}
In this work we consider a \emph{power} probability weighting function as $\pi(p)\coloneqq p^S$, $S\in \mathds{R}$ \cite{PT_review_2013, PT_review_2010, PT_review_2004, PT_review_2000}. This function encapsulates the behaviours of optimism $(S<1)$ as well as pessimism $(S>1)$ \cite{mohammed2010, weber1994}. We now establish our main results.

\section{Main Results}
\label{s:R}

We now start with the presentation of our main results. We first introduce operational tasks based on betting for the utility theory of \emph{wealth}, where both gambler and bookmaker have access to side information, or \emph{double} side information for short. We then characterise these tasks in terms of information-theoretic quantities and generalise these results to prospect theory. We then introduce a utility theory for \emph{wealth-ratios} and interpret there the Ili\'c-Djordjevi\'c mutual information measure. We then also derive a generalised chain rule for the Ili\'c-Djordjevi\'c conditional entropy. Finally, we address the implications of these results on the framework of quantum resource theories.

\subsection{Utility theory of wealth and betting games with double side information}
\label{ss:BT}

We invoke a utility theory of \emph{wealth} and address here operational tasks involving betting. The utility theory of \emph{wealth} represents the quintessential example of a utility theory, since it is the default utility that is considered in the economic sciences. In the utility theory of wealth, we can address \emph{decision problems} where a rational agent is asked to make a choice between two lotteries, and these decisions are going to give away the type of utility function that represents the agent's economic behaviour. These decision problems, whilst useful in this regard, are at the same time somewhat limited, in the sense that the actions of rational agents are effectively limited to (only) \emph{decide} between various options and, in this sense, the agent can be seen as playing a \emph{passive} role in the scenario. In this regard, there is nevertheless a natural way to allow the rational agent to have a more \emph{active} participation  in the decision problem, the idea being to allow the rational agent to implement \emph{betting}. In a betting scenario, as in a decision problem, the rational agent is still presented with a set of options to choose from, but now the agent is additionally going to be allowed to \emph{place bets} on the lotteries in question, so effectively have a more active role. Let us now address this more formally.

A general \emph{betting task} is a protocol involving two agents: a referee and a gambler. Consider a random event distributed according to the PMF $p_X$. Similar to a decision problem, the referee is going to ask the gambler to decide between a lottery or a fixed (certain) amount of wealth. The lottery in question is given by a pair $(p_X,w_X)$, with $w_X$ a distribution of wealth. The main difference with a standard decision problem is that now the gambler can play a more active role when it comes to the amount of wealth that he is going to receive. This is because the wealth to be rewarded is now considered a function with two components as $w_X=b_Xo_X$, with $o_X$ a function refereed to as \emph{the odds} (not necessarily a PMF), which is proposed by the referee, and a PMF $b_X$, which represents \emph{the bet} placed by the gambler. The idea is that after correctly guessing the outcome of the random event, the referee rewards the gambler with an amount of wealth given by $w(x)=b(x)o(x)$. It is then explicit that the gambler can influence the amount of wealth to be rewarded. In this decision problem involving betting, or \emph{betting task} for short, the quantity that describes the gambler's assessment of the lottery $(p_X, w_X)$ is the certainty equivalent of wealth given by:
\begin{align}
    w^{\rm CE}_u
    (b_{X},o_{X},p_{X})
    =
    u^{-1}
    \left(
    \mathbb{E}_{p_{X}}[
    u
    (
    w_{X}
    \coloneqq
    b_X
    o_X    
    )
    ]
    \right)
    ,
\end{align}
which represents the minimum amount of wealth that the gambler is willing to accept so to walk away from the lottery. We can similarly introduce scenarios dealing with negative wealth, where the decision problem and consequently the betting task is now dealing with \emph{losses} instead of gains. This scenario can naturally be included by allowing the odds function to be negative (full details about this construction in \cite{DS2022}).

One common characteristic amongst works in the literature addressing betting tasks so far, is the assumption of \emph{side information} being available \emph{exclusively} to the gambler \cite{kelly, BLP1, DS2022}. In this work we further extend this to include a more elaborate scenario in which \emph{both} gambler and bookmaker can now have access to side information. We coin this scenario as betting tasks with \emph{double} side information, and formalise it as follows.

\begin{definition}(Betting tasks with double side information)
    Consider a random event described by the PMF $p_X$. Consider also random variables $G$ (gambler's side information) and $Y$ (bookmaker's side information)  correlated with the random variable $X$ and described by the joint PMF $p_{XGY}$. Consider now a game (betting task) with a referee and a gambler (rational agent), with the latter being represented by an isoelastic utility function $u_R^{\rm I}(\cdot)$ with a constant relative risk aversion (CRRA) factor $R\in \mathds{\overline R}$. The game consists on the gambler placing ``a bet" with the help of side information $G$ as $b_{X|G}$ (a conditional PMF) on the output of the random event. The referee then rewards the gambler (when guessing correctly) with an amount of wealth given by $w(x,g,y) = b(x|g)\,o(x|y)$, with $o_{X|Y}$ the conditional ``odds", a pre-established function proposed by the referee (not necessarily a PMF), and known to the gambler. The figure of merit of interest here is the isoelastic certainty equivalent (ICE), which for risk values $R \in (-\infty,1) \cup (1,\infty)$ can be written as:
    \begin{align}
    	&
    	w^{\rm ICE}_R
    	(b_{X|G},o_{X|Y},p_{XGY})
    	\nonumber
    	\\
    	&=
    	(u_R^{\rm I})^{-1}
        \left(
        \mathbb{E}_{p_{XGY}}[
        u_R^{\rm I}
        (w_{XGY})
        ]
        \right)
    	\\
    	&=
    	\left[
    	\sum_{g,x,y}
    	\big[
    	b(x|g)
    	\,
    	o(x|y)\big]^{1-R}
    	p(x,g,y)
    	\right]^\frac{1}{1-R}
    	\hspace{-0.5cm}
    	.
    	\label{eq:HB_SI_W}
    \end{align}
    The cases $R \in\{1, \infty, -\infty\}$ are defined by continuous extension of \eqref{eq:HB_SI_W}. A betting task is specified by $(o_{X|Y}, p_{GXY})$, and the gambler plays this game with a betting strategy $b_{X|G}$.
\end{definition}
We now address some particular cases of interest.
\begin{remark} (Betting tasks with individual side information)
    It is useful to specify here two natural scenarios. First, a scenario where only the gambler has access to side information (or betting task with gambler's side information):
    {\small\begin{align}
    	w^{\rm ICE}_R
    	(b_{X|G},o_X,p_{XG})
    	=
    	\left[
    	\sum_{g,x}
    	\big[
    	b(x|g)
    	\,
    	o(x)\big]^{1-R}
    	p(x,g)
    	\right]^\frac{1}{1-R}
    	\hspace{-0.5cm}
    	.
    	\label{eq:BT_SIG}
    \end{align}}
    Second, a scenario where only the bookmaker has access to side information (or betting task with bookmaker's side information):
    {\small\begin{align}
    	w^{\rm ICE}_R
    	(b_{X},o_{X|Y},p_{XY})
    	=
    	\left[
    	\sum_{y,x}
    	\big[
    	b(x)
    	\,
    	o(x|y)\big]^{1-R}
    	p(x,y)
    	\right]^\frac{1}{1-R}
    	\hspace{-0.5cm}
    	.
    	\label{eq:BT_SIB}
    \end{align}}
\end{remark}
One useful concept to introduce here is that of the \emph{fairness of the odds} \cite{CT, BLP1}. Given a conditional odds function $o_{X|Y}$, we consider the quantity:
\begin{align}
    c^{(o)}
    (y)
    \coloneqq
    \left(
    \sum_x 
        |o(x|y)|^{-1}
    \right)^{-1}
    ,
\end{align}
and classify the fairness of the odds as follows:
\begin{align}
    c^{(o)}
    (y)
    &
    =1,
    \forall y,
    \hspace{0.5cm}\text{Fair odds}
    .
    \\
    c^{(o)}
    (y)
    &>
    1,
    \forall y,
    \hspace{0.5cm}\text{Superfair odds},
    \\
    c^{(o)}
    (y)
    &<
    1,
    \forall y,
    \hspace{0.5cm}\text{Subfair odds}
    .
\end{align}
Finally, we allow the gambler to optimise over all possible betting strategies, for a given game setup $(o_{X|Y}, p_{XGY})$:
\begin{align}
    \max_{b_{X|G}}
    w^{\rm ICE}_R
	(b_{X|G},o_{X|Y},p_{XGY})
	,
\end{align}
with the maximisation over all conditional PMFs. It is also going to be useful to introduce the following auxiliary function, which we here simply address as the logarithm of the ICE:
{\small\begin{align}
    &U^{\rm ICE}_R
	(b_{X|G},o_{X|Y},p_{XGY})
	\\
	&\coloneqq
	\sgn(o)
    \ln
     \left[
     w^{\rm ICE}_{R}
     (b_{X|G}, |o_{X|Y}|, p_{XGY})
     \right]
     .
     \nonumber
\end{align}}
We are now ready to address our main results concerning the characterisation of these operational tasks in terms of information-theoretic quantities.

\subsection{Information-theoretic characterisation of betting games with double side information}
\label{ss:BT2}

In this subsection we address the characterisation of various types of betting tasks in terms of information-theoretic quantities. We start with betting tasks with bookmaker's side information (only the bookmaker has access to side information), followed by general betting tasks with \emph{double} side information (both gambler and bookmaker have access to side information). We compare the performance of a rational agent when playing betting tasks where there is bookmaker's side information, against a scenario with no side information at all. Similarly, we then compare the performance of a rational agent when playing betting tasks where there is bookmaker's side information against a scenario with gambler's side information. 

\begin{result} \label{r:r3}
(Characterisation of betting games with  bookmaker's side information)
    Consider a betting game defined by the pair $(o_{X|Y}, p_{XY})$, with fair odds as $c^{(o)}(y)=\sum_x |o(x|y)|^{-1}=1$, $\forall y$. Consider now a gambler playing this game with a betting strategy given by $b_X$. Then, we have that the logarithm of the ICE is characterised by the n1-CR divergence $D_\alpha^{\rm n1}(\cdot||\cdot|\cdot)$ and the R-divergence $D_\alpha(\cdot||\cdot)$ as:
	\begin{align}
    	&
    	U^{\rm ICE}_{R}
         \left(
         b_{X}
         ,
         o_{X|Y}
         , 
         p_{XY}
         \right)
         \\
    	&=
    	\sgn(o)
    	\sgn(R)\,
    	D_{1/R}^{
    	\rm n1
    	}
    	(
    	p_{X|Y}
    	||
    	r^{(o)}_{X|Y}
    	|
    	p_Y
    	)
    	\nonumber
    	\\
    	&
    	-
    	\sgn(o)
    	\sgn(R)\,
    	D_{R}
    	(h^{(R,o,p)}_X||b_X)
    	,
    	\nonumber
	\end{align}
	with the PMFs $r^{(o)}(x|y)
	\coloneqq
	|o(x|y)|^{-1}$, 
	and:
	\begin{align}
		h^{(R,o,p)}
		(x)
		&\coloneqq 
		\frac{
		    \left(
    		    \sum_y
    			p(xy)
    			\,
    			|o(x|y)|^{1-R}
			\right)^\frac{1}{R}
		}
		{
			\sum_{x'}
			\left(
    		    \sum_y
    			p(x'y)
    			\,
    			|o(x'|y)|^{1-R}
			\right)^\frac{1}{R}
		}
		.
		\end{align}
\end{result}
The proof of this result is in \cref{a:r3}. This result characterises betting tasks with bookmaker's side information. In particular, a consequence of this result is the explicit form of the optimal betting strategy for the gambler to play these games, as well as an operational interpretation of the n1-CR divergence.
\begin{corollary} \label{c:c1} (Optimal betting strategy for betting tasks with bookmaker's side information)
    Consider a betting scenario with $\sgn(o)=\sgn(R)$. We can readily check, from the previous decomposition, that the optimal betting strategy is given by $b_X^*=h_X^{(R,o,p)}$, which achieves the quantity:
    {\small\begin{align}
        \max_{b_{X}}
        U^{\rm ICE}_R
    	(b_{X},o_{X|Y},p_{XY})
    	=
    	D_{1/R}^{
    	\rm n1
    	}
    	(
    	p_{X|Y}
    	||
    	r^{(o)}_{X|Y}
    	|
    	p_Y
    	)
    	.
    \end{align}}
\end{corollary}
We can also recover the known case where there is no side information as a corollary \cite{BLP1}.
\begin{corollary}\label{c:c2} 
(Optimal betting strategy for betting tasks with no side information)
    Consider a betting scenario with $\sgn(o)=\sgn(R)$. From the previous corollary, and using the fact that the n1-CR-divergence reduces to the R\'enyi divergence when $X \indep Y$, we have:
    \begin{align}
        \max_{b_{X}}
        U^{\rm ICE}_R
    	(b_{X},o_{X},p_{X})
    	=
    	D_{1/R}
    	(
    	p_{X}
    	||
    	r^{(o)}_{X}
    	)
    	.
    \end{align}
\end{corollary}
With these two corollaries in place, we can now analyse the performance of a rational agent when playing a betting task with bookmaker's side information in comparison to a scenario with no side information.

\begin{corollary} \label{r:c4}
    (Ratio of betting games between bookmaker's side-information and no side-information)
    Consider $R\in[1,+\infty]$, with the odds function for each betting game related as $o(x)\coloneqq \sum_y o(x|y)p(y)$, then:
	\begin{align}
    	&
        \ln
    	\left[
    	\frac{
    		\displaystyle
    		\max_{b_{X}}
    		\,
    		w^{\rm ICE}_{R}
    		(
    		b_{X}
    		,
    		o_{X|Y}
    		,
    		p_{XY}
    		)
    	}{
    		\displaystyle
    		\max_{b_{X}}
    		\,
    		w^{\rm ICE}_{R}
    		(
    		b_X
    		,
    		o_X
    		,
    		p_X
    		)
    	}
    	\right]
    	\\
    	&=
    	D_{1/R}^{
    	\rm n1
    	}
    	(
    	p_{X|Y}
    	||
    	r^{(o)}_{X|Y}
    	|
    	p_Y
    	)
    	-
    	D_{1/R}
    	(
    	p_X
    	||
    	r_X^{(o)}
    	)
        \geq 
        0
        ,
        \nonumber
    \end{align}
    with
    the PMFs $p_X$, $r_X$ given by 
    $
     p(x)
    \coloneqq
    \sum_y
    p(x|y) 
    p(y)
    $,
    $
    r(x)
    \coloneqq
    \sum_y
    r(x|y) 
    p(y)
    $.
\end{corollary}
Details about this corollary in \cref{a:c4}. The inequality in the corollary follows from the data processing inequality derived in \cref{r:r1}. This corollary is telling us that it is \emph{more} useful for the gambler when the bookmaker has access to side information, as opposed to no side-information at all. This is a rather counterintuitive statement, because one can naively expect that it should actually be \emph{worse} for the gambler, when the bookmaker is the one having access to side information. This is however, as the maths is telling us, not the case. The explanation of this seemingly paradoxical situation is that, even though the bookmaker is the agent explicitly having access to the side information, we should not forget that she is still in charge of proposing the conditional odds function $o_{X|Y}$, which itself depends on $Y$, and that this odds in turn are going to be known to the gambler (as per the rules of the betting game), so that ultimately the gambler \emph{does} have access to the side information $Y$, although via this implicit manner. From a technical point of view, it is also interesting that the inequality holds for $R\geq 1$, and so it could prove insightful to understand its behaviour for other values, but we leave this however for future research. We now can also compare rational agents playing betting tasks with bookmaker's side information and gambler's side information.

\begin{corollary} \label{r:c5}
    (Ratio of betting games between bookmaker's side-information and gambler's side-information)
    Consider $R\in \mathds{\overline R}$ then:
	\begin{align}
    	&\sgn(R)
            \ln
    	\left[
    	\frac{
    		\displaystyle
    		\max_{b_{X|G}}
    		\,
    		w^{\rm ICE}_{R}
    		(
    		b_{X|G}
    		,
    		o_{X}
    		,
    		p_{XG}
    		)
    	}{
    		\displaystyle
    		\max_{b_{X}}
    		\,
    		w^{\rm ICE}_{R}
    		(
    		b_{X}
    		,
    		o_{X|G}
    		,
    		p_{XG}
    		)
    	}
    	\right]
    	\\
    	&=
    	D_{1/R}^{
    	\rm BLP
    	}
    	(
    	p_{X|G}
    	||
    	r^{(o)}_{X|G}
    	|
    	p_G
    	)
    	-
    	D_{1/R}^{
    	\rm n1
    	}
    	(
    	p_{X|G}
    	||
    	r^{(o)}_{X|G}
    	|
    	p_G
    	)
        \geq 0
        ,
        \nonumber
    \end{align}
    with the odds function for each betting game related as $o(x)\coloneqq \sum_g o(x|g)p(g)$.
\end{corollary}
Details about this corollary in \cref{a:c5}. This corollary is telling us that side-information is \emph{more useful} when given to the gambler directly, as opposed to when it is given to the bookmaker. This confirms what one can intuitively expect, with the added benefit that this corollary allows us to \emph{quantify} the advantage provided by giving the side information \emph{directly} to the gambler. We now move on to the more general scenario of betting tasks with \emph{double} side information.
\begin{result}\label{r:r4}
(Characterisation of betting games with double side information)
    Consider a betting game defined by the pair $(o_{X|Y},p_{XGY})$, and fair odds so that $c^{(o)}(y)=\sum_x |o(x|y)|^{-1}=1$, $\forall y$. Consider also a gambler playing this game with a betting strategy $b_{X|G}$. Then, the logarithm of the ICE is characterised by the the n2-CR-divergence 
    $
    D^{\rm n2}_\alpha
    (
     \cdot
     ||
     \cdot
     |
     \cdot
     )$ and the R-divergence $D_\alpha(\cdot||\cdot)$ as:
    \begin{align}
		&U_R
		(b_{X|G},o_{X|Y},p_{XGY})
		\\
		&=
		\sgn(o)
		\sgn(R)
		D^{\rm n2}_{1/R}
		\left(
    		p_{X|GY}
    		||
    		r^{(o)}_{X|Y}
    		|
    		p_{G}
    		,
    		p_{Y|G}
		\right)
		\nonumber
		\\
		&-
		\sgn(o)
		\sgn(R)
		D_{R}
		\left(
    		h^{(R,o,p)}_{X|G} 
    		h^{(R,o,p)}_{G}
    		\Big| \Big|
    		b_{X|G}\,
    		h^{(R,o,p)}_{G}
		\right),
		\nonumber
	\end{align}
	with the conditional PMF $r^{(o)}(x|y)
	\coloneqq
	|o(x|y)|^{-1}$, 
	and the conditional PMF and PMF:
	\small{
	\begin{align}
		h^{(R,o,p)}
		(x|g)
		&\coloneqq 
		\frac{
		    \left(
    		    \sum_y
    			p(xy|g)
    			\,
    			|o(x|y)|^{1-R}
			\right)^\frac{1}{R}
		}
		{
			\sum_{x'}
			\left(
    		    \sum_y
    			p(x'y|g)
    			\,
    			|o(x'|y)|^{1-R}
			\right)^\frac{1}{R}
		},
		\label{eq:HB_SI_b}
		\\
		h^{(R,o,p)}
		(g)
		&\coloneqq \frac{
		    p(g)
		    \left[
		    \sum_{x}
    			\left(
        		    \sum_y
        			p(xy|g)
        			\,
        			|o(x|y)|^{1-R}
			    \right)^\frac{1}{R}
			\right]^{R}
		}
		{
			\sum_{g'}
			p(g')
		    \left[
		    \sum_{x}
    			\left(
        		    \sum_y
        			p(xy|g')
        			\,
        			|o(x|y)|^{1-R}
			    \right)^\frac{1}{R}
			\right]^{R}
		}.
		\nonumber
		\end{align}}
\end{result}
The proof of this result is in \cref{a:r4}. Similar to the previous case, we can also identify here the optimal betting strategy to be implemented by a gambler playing these games, and provide an operational interpretation of the n2-CR divergence.
\begin{corollary} \label{c:c3} 
(Optimal betting strategy for betting tasks with double side information)
    Consider a betting scenario with $\sgn(o)=\sgn(R)$. We can readily check, from the previous decomposition, that the optimal betting strategy is given by $b_{X|G}^*=h_{X|G}^{(R,o,p)}$, which achieves the quantity:
    \begin{align}
        &\max_{b_{X|G}}
        U^{\rm ICE}_R
    	(b_{X|G},o_{X|Y},p_{XGY})
    	\\
    	&=
    	D_{1/R}^{
    	\rm n2
    	}
    	(
    	p_{X|GY}
    	||
    	r^{(o)}_{X|Y}
    	|
            p_G,
    	p_{Y|G}
    	)
    	.
    	\nonumber
    \end{align}
\end{corollary}
From this general result, and using the relationship between conditional divergences, we can recover the previous corollaries as particular cases. We now move on to further generalise these results, from the theory of expected utility to prospect theory.

\subsection{Generalisation to original prospect theory}
\label{ss:PT}

In this section we address betting tasks from the point of view of prospect theory. In particular, we analyse here betting tasks with no side information, as well as betting tasks with gambler's side information. Additionally, we consider here the certainty equivalent for prospect theory with the \emph{isoelastic} utility function together with the  \emph{power} probability weighting function.  

\begin{result} \label{r:r5} (Characterisation of betting tasks with no side information within prospect theory)
    Consider a betting game specified by the pair $(o_X, p_X)$, and a gambler represented by the pair $(R,S)$ ($R$ is the standard CRRA whilst $S$ is the coefficient related to the agent's sensitivity when assessing probabilities) playing this game with a betting strategy $b_X$. The logarithm of the ICE within prospect theory is characterised by the R-divergence $D_\alpha(\cdot||\cdot)$ and the R\'enyi entropy $H_\alpha(\cdot)$ as:
    \begin{align}
    &
    \sgn(o)
    \ln
     \left[
         w^{\rm CE-PT}_{R, S}
         (
         b_X
         ,
         |o_X|
         ,
         p_X
         )
     \right]
     \\
    	&=
    	\sgn(o)
    	\sgn(S)
    	\frac{1-S}{1-R}
    	H_S
    	(p_X)
    	+
    	\sgn(o)
    	\log 
    	|
    	    c^{(o)}
    	|
    	+
    	\nonumber
    	\\
    	&+
    	\sgn(o)
    	\sgn(R)\,
    	D_{1/R}
    	(q_X^{(S,p)}||r_X^{(o)})
    	\nonumber
    	\\
    	&-
    	\sgn(o)
    	\sgn(R)\,
    	D_{R}
    	(h^{(R, S,o,p)}_X||b_X)
    	,
    	\nonumber
	\end{align}
	with the parameter and the PMFs:
	\begin{align}
    	&c^{(o)}
    	\coloneqq
    	\left(
    	\sum_x 
    	o(x)^{-1}
    	\right)^{-1}
    	,
    	\hspace{0.5cm}
    	    r^{(o)}
    	    (x)
        	\coloneqq
            	\frac{
            	c^{(o)}
            	}{
            	o(x)
            	}
        	,
        \\
        &h^{(R, S,o,p)}(x)
    	\coloneqq
    	\frac{
    		q(x)^{\frac{1}{R}}
    		o(x)^{\frac{1-R}{R}}
    	}
    	{
    		\sum_{x'}
    		q(x')^{\frac{1}{R}}
    		o(x')^{\frac{1-R}{R}}
    	}
    	,
    	\\
    	&q^{(S,p)}(x)
    	\coloneqq
    	\frac{
    	p(x)^S
    	}{
    	\sum_{x'}
    	p(x')^S
    	}
            .
	\end{align}
	The functions $r_X^{(o)}$ and $h_X^{(R,S,o,p)}$ define valid PMFs even for negative odds ($o(x)<0$, $\forall x$).
\end{result}
The proof of this result is in \cref{a:r5}. We now consider a CE of wealth with the \emph{isoelastic} utility function and the \emph{power} probability weighting function in a scenario with gambler's side information. We find the following characterisation.

\begin{result} \label{r:r6} (Characterisation of betting tasks with gambler's side information within prospect theory)
    Consider a betting game specified by the pair $(o_X, p_{XG})$, and a gambler represented by the pair $(R,S)$ playing this game with a betting strategy $b_{X|G}$. The logarithm of the ICE is characterised by the R-divergence $D_\alpha( \cdot || \cdot)$, the BLP-CR divergence $D^{\rm BLP}_\alpha( \cdot || \cdot)$, and the R\'enyi entropy $H_\alpha(\cdot)$ as:
    \begin{align}
    &
    \sgn(o)
    \ln
     \left[
         w^{\rm CE-PT}_{R, S}
         (
         b_{X|G}
         ,
         |o_X|
         ,
         p_{XG}
         )
     \right]
     \\
     &=
     \sgn(o)
     \sgn(S)
    	\frac{1-S}{1-R}
         H_S
         (p_{XG})
          +
        \sgn(o)
    	\log 
    	|
    	    c^{(o)}
    	|
    	+
    	\nonumber
    	\\
    	&
        +
    	\sgn(o)
    	\sgn(R)\,
    	D_{1/R}^{\rm BLP}
    	(
    	q_{X|G}^{(S,p)}
    	||
    	r^{(o)}_X
    	|
    	q_G^{(S,p)})
    	+
    	\nonumber
    	\\
    	&-
    	\sgn(o)
    	\sgn(R)\,
        D_{R}
		(
		h^{(R, S, o, p)}_{X|G}
		h^{(R, S, o, p)}_G
		||
		b_{X|G}
		h^{(R, S, o, p)}_G
		)
    	,
    	\nonumber
	\end{align}
	with the parameter and the PMF:
	\begin{align}
    	c^{(o)}
    	\coloneqq
    	\left(
    	\sum_x o(x)^{-1}
    	\right)^{-1},
    	\hspace{0.5cm}
    	    r^{(o)}
    	    (x)
        	\coloneqq
            	\frac{
            	c^{(o)}
            	}{
            	o(x)
            	}
        	,
	\end{align}
	and the conditional PMF, PMF, and escort PMF:
	{\small\begin{align}
		h^{(R, S,o,p)}(x|g)
		&\coloneqq \frac{
			q(x|g)^{\frac{1}{R}}
			o(x)^{\frac{1-R}{R}}
		}
		{
			\sum_{x'}
			q(x'|g)^{\frac{1}{R}}
			o(x')^{\frac{1-R}{R}}
		},
		\label{eq:HB_SI_b_prospect}
		\\
		h^{(R, S,o,p)}(g)
		&\coloneqq \frac{
		    p(g)
		    \left[
		    \displaystyle
		    \sum_{x'}
			q(x'|g)^{\frac{1}{R}}
			o(x')^{\frac{1-R}{R}}
			\right]^{R}
		}
		{
		    \displaystyle
			\sum_{g'}
			p(g')
		    \left[
		    \displaystyle
		    \sum_{x'}
			q(x'|g')^{\frac{1}{R}}
			o(x')^{\frac{1-R}{R}}
			\right]^{R}
		},
		\\
		q^{(S,p)}(x|g)
		&\coloneqq
		\frac{
		p(x|g)^S
		}{
		\sum_{x'}
		p(x'|g)^S
		}
		,
		\\
		q^{(S,p)}(g)
		&\coloneqq
		\sum_x
		q^{(S,p)}(x,g)
		.
		\end{align}}
		The quantities $r_X^{(o)}$, $h^{(R, S,o,p)}_{X|G}$, $h^{(R, S,o,p)}_G$ define valid PMFs even for negative odds ($o(x)<0$, $\forall x$).
\end{result}
The proof of this result is in \cref{a:r6}. We now consider constant odds as $o^{\sgn(\alpha)c}(x) \coloneqq \sgn(\alpha)C$, $C>0$, $\forall x$, and compare the two previous results to quantity the advantage provided by side information.

\begin{corollary} (Advantage provided by gambler's side information for betting tasks within prospect theory)
    Consider a betting game defined by $(o^{\sgn(R)c}_X, p_{XG})$ with constant odds as $o^{\sgn(\alpha)c}(x) \coloneqq \sgn(\alpha)C$, $C>0$, $\forall x$, and a joint PMF $p_{XG}$. Consider a first gambler having access to side information $G$, against a second gambler with no access to side information. Both gamblers are represented by isoelastic functions $u_R^{\rm I}(\cdot)$. The gamblers are allowed to independently optimise overall betting strategies. We have the following relationship:
	\begin{align}
    	&\sgn(R)
        \ln
    	\left[
    	\frac{
    		\displaystyle
    		\max_{b_{X|G}}
    		\,
    		w^{\rm CE-PT}_{R, S}
    		(
    		b_{X|G}
    		,
    		o^{\sgn(R)c}_X
    		,
    		p_{XG}
    		)
    	}{
    		\displaystyle
    		\max_{b_{X}}
    		\,
    		w^{\rm CE-PT}_{R, S}
    		(
    		b_X
    		,
    		o^{\sgn(R)c}_X
    		,
    		p_X
    		)
    	}
    	\right]
    	\\
    	&=
    	I_{1/R}^{\rm A}
    	(
    	q_{XG}^{(S,p)}
    	)
    	+
    	\sgn(RS)
    	\frac{1-S}{1-R}
        H_S^{2}
        (p_{G|X}p_X)
        ,
        \nonumber
    \end{align}
    with Arimoto's mutual information and the second R\'enyi conditional entropy.
\end{corollary}
We note that reducing to standard expected utility theory (by specifying $S=1$), the second terms vanishes, and we recover the Result 5 in \cite{DS2022}. This provides an operational interpretation for the second R\'enyi conditional entropy; it quantifies the deviation achieved when considering prospect theory against standard expected utility theory. Finally, we note that we can similarly define variants for betting tasks with bookmaker's side information, and double side information and derive similar decompositions. This finishes our main results concerning the utility theory of wealth. We now consider a utility theory of wealth ratios.

\subsection{Utility theory of wealth ratios}
\label{ss:WR} 

We now introduce an extension of the utility theory of wealth. Let us start with the standard setup for a utility theory of wealth. Consider a PMF $p_X$ describing a random event, a rational agent represented by the utility function $u(\cdot)$, and the scenario for a betting task so that we have the rational agent's CE as 
$w^{\rm CE}_{u}
\left(
b_{X},
o_X,
p_{X}
\right)$. The motivation now is to compare the performance of different betting strategies for the same betting task $(o_X, p_X)$, this means that we want to consider two different betting strategies $b_X^1$ and $b_X^2$ whilst keeping the same gambler (same utility function $u(\cdot)$), same odds function $o_X$, the same PMF $p_X$ . Explicitly, this amounts to compare the two quantities:
\begin{align}
    w^{\rm CE}_{u}
    \left(
    b_{X}^1,
    o_X,
    p_{X}
    \right)
        ,
    \hspace{0.5cm}
    w^{\rm CE}_{u}
    \left(
    b_X^2,
    o_X,
    p_X
    \right)
    .
\end{align}
Without of loss of generality, we can assume that one of them (say the first one) is greater or equal than the other (say the second one) and define the \emph{advantage} that betting strategy $b_X^1$ has over betting strategy $b_X^2$ as:
\begin{align}
    A^{1,2}_u
    \left(
    b_X^1
    ,
    b_X^2
    ,
    o_X
    ,
    p_X
    \right)
    \coloneqq
    \frac{
        w^{\rm CE}_{u}
    	\left(
    	b_{X}^1,
    	o_X,
    	p_{X}
    	\right)
    }{
        w^{\rm CE}_{u}
        \left(
        b_X^2,
        o_X,
        p_X
        \right)
    }
    \geq 
    1
    .
    \label{eq:A2}
\end{align}
This ratio is operational in nature, as it is explicitly comparing the performance of the betting strategies $b_X^1$ and $b_X^2$. This ratio can then be understood as the \emph{wealth ratio} or the \emph{advantage} that the first betting strategy offers against the second betting strategy. We also note that $A_u^{1,2}\in [1, \infty)$. In a similar manner, and when considering scenarios with gambler's side information, a natural ratio that emerges is the following:
\begin{align}
    A^{\rm SI}_u
    (
    o_X
    ,
    p_{XG}
    )
    \coloneqq
    \frac{
		\displaystyle
		\max_{b_{X|G}}
		\,
		w^{\rm CE}_u
		(
		b_{X|G},o_X,p_{XG}
		)
	}{
		\displaystyle
		\max_{b_{X}}
		\,
		w^{\rm CE}_u
		(
		b_X,o_X,p_X
		)
	}
	,
	\label{eq:ASI}
\end{align}
with $p(x)\coloneqq \sum_g p(x|g)p(g)$. In this case, the wealth ratio in question is quantifying the advantage provided by side information $G$. Given these two ratios, it then seems natural to consider a utility theory based on ``advantage", given in terms of \emph{wealth ratios}, instead of \emph{wealth} by itself. Now, in general terms, we can naturally address a new utility theory based on such \emph{wealth ratios}, which can be interpreted as the level of satisfaction that a rational agent would experience with the prospect of achieving such an advantage. The quantities which naturally arise in such utility theory based on wealth ratios are then given by:
\begin{align}
    &v
    \left(
    A^{1,2}_u
        \left(
        b_X^1
        ,
        b_X^2
        ,
        o_X
        ,
        p_X
        \right)
    \right)
    ,
    \\
    &v
    \left(
    A^{\rm SI}_u
        \left(
        o_X
        ,
        p_{XG}
        \right)
    \right)
    ,
\end{align}
with $v:[1, \infty)\rightarrow \mathds{R}$ a utility function for \emph{wealth ratios}, and $u(\cdot)$ a standard utility function for \emph{wealth}. We note that these two utility functions can be different, as rational agents can have a specific attitude with respect to wealth, and a different when considering the growth of such wealth. Whilst this formalism can be introduced for general utility functions $(u,v)$, for simplicity, let us assume rational agents described by isoelastic utility functions as $u(\cdot)=u_R^{\rm I}(\cdot)$, $v(\cdot)=u_S^{\rm I}(\cdot)$, with different attitude to risk as $S\neq R$. We are now ready to describe betting tasks from the point of view of utility theories of wealth ratios.

\subsection{Sharma-Mittal information-theoretic quantities and the EUT of wealth ratios}
\label{ss:WR2}

In this section we characterise the advantage provided by side information within utility theories of wealth ratios. We then derive a generalised chain rule for the ID conditional entropy.

\begin{result} \label{r:r7} 
(Operational interpretation of the Ili\'c-Djordjevi\'c mutual information measure)
The Ili\'c-Djordjevi\'c mutual information measure $I_{q,r}^{\rm ID}(X;G)$ of orders $(q,r)\in \mathds{\overline R} \times \mathds{R}$ of a joint PMF $p_{XG}$ is given by:
\begin{align}
    &I^{\rm ID}_{q,r}(X;G)
    \\
    &=
    \sgn(q)\,
    u_r
    \left[
        \frac{
		\displaystyle
		\max_{b_{X|G}}
		\,
		w^{\rm ICE}_{1/q}
		(
		b_{X|G},
		o_X^{\sgn(q)c},
		p_{XG}
		)
	    }{
		\displaystyle
		\max_{b_{X}}
		\,
		w^{\rm ICE}_{1/q}
		(
		b_X,
		o_X^{\sgn(q)c},
		p_X
		)
	    }
    \right]
    .
    \nonumber
\end{align}
The Ili\'c-Djordjevi\'c entropy quantifies the utility of a $r$-agent on the wealth ratio (advantage) provided by side information $G$ achieved by a $1/q$-agent when the latter plays betting tasks where odds are constant.
\end{result}
The proof of this result is in \cref{a:r7}. This result holds true for any joint PMF $p_{XG}$ as well as for any combination of orders $(q,r)\in \mathds{\overline R} \times \mathds{R}$. Finally, due to the generality of this result, we can derive operational interpretations for the generalised mutual information measures of R\'enyi, Tsallis, Gauss, when taking the respective limits. We now derive a generalised chain rule for the Ili\'c-Djordjevi\'c conditional entropy.

\begin{result} \label{r:r8}
(Generalised chain rule for the Sharma-Mittal entropy and the Ili\'c-Djordjevi\'c conditional entropy) 
    The Ili\'c-Djordjevi\'c conditional entropy of orders $(q,r)\in\mathds{\overline R} \times \mathds{R}$ satisfies the condition:
    {\small\begin{align}
        \sgn(q)
        H^{\rm ID}_{q,r}(X|G)
        \geq
        \sgn(q)
        \left[
            H^{\rm SM}_{q,r}(XG)
            \ominus_r
            \ln_r
            \left(
            K
            \right)
        \right]
        ,
    \end{align}}
    with $\ominus_r$ the pseudo-substraction. 
\end{result}
The proof of this result is in \cref{a:r8}. In particular, we recover a known case in the literature as a limit case.
\begin{corollary} 
    We recover the case for the Arimoto-R\'enyi conditional entropy. For $r=1$, $q\geq0$, $H^{\rm A}_{q}(X|G) \geq H^{\rm R}_{q}(X)-\log(K)$ \cite{review_RCE}.
\end{corollary}
Considering this property, together with the operational interpretation in terms of betting tasks, the Ili\'c-Djordjevi\'c conditional entropy stands out as a very appealing conditional entropy \`a la Sharma-Mittal. In addition to this, it is also worth highlighting that its associated mutual information satisfies various desirable properties, which are not simultaneously valid for other alternatives \`a la Sharma-Mittal \cite{ID1}. Altogether, these results therefore make the quantities introduced by Ili\'c-Djordjevi\'c stand as strong generalisations of  entropic quantities \`a la Sharma-Mittal. We now address the implications of these latter results on the field of quantum resource theories.

\subsection{Implications on the QRTs of informative measurements and non-constant channels}
\label{ss:R45}

We now consider the implications of the previous results on quantum resource theories, specifically, the QRTs of measurement informativeness and non-constant channels. A \emph{quantum state} is a trace one ($\tr(\rho)=1$) positive semidefinite operator ($\rho\geq0$) acting on a finite-dimensional Hilbert space. A positive operator-valued measure (POVM) or \emph{quantum measurement} is a set of positive semidefinite operators $\mathbb{M}=\{M_a\geq0\}$ such that $\sum_a M_a = \mathds{1}$. An \emph{ensemble of quantum states} is a set $\mathcal{E} = \{\rho_x, p(x)\}$, $x\in\{1,...,K\}$, with $p_X$ a PMF. A \emph{quantum channel} is a completely positive  trace-preserving (CPTP) map \cite{NC}.

In the framework of quantum resource theories (QRTs), it is common to first specify the mathematical \emph{objects} of the theory, followed by a \emph{property} of such objects to be considered as a \emph{resource} \cite{RT_review}. We invoke here a QRT of measurements with the resource of \emph{informativeness} \cite{SL}. A POVM $\mathbb{M}=\{M_g\}$ is called \emph{uninformative} when there exists a PMF $q_G$ such that the POVM effects can be written as $M_g = q(g)\mathds{1}$, $\forall g$ \cite{SL, DS}. The set of uninformative measurements is denoted as $\rm UI$. We also invoke here a QRT of channels with the resourceful channels being \emph{non-constant} channels. A quantum channel $\mathcal{N}$ is called \emph{constant} when there exists a state $\sigma_\mathcal{N}$ such that $\mathcal{N} (\rho) = \sigma_\mathcal{N}$, $\forall \rho$. The set of constant channels is denoted as $\mathcal{C}$. There are two natural betting tasks to be considered in the quantum domain, quantum state betting (QSB) and noisy quantum state betting (NQSB) \cite{DS2022}.

One pragmatical way of introducing quantum state betting (QSB) games is to consider a standard betting game where the conditional PMF $p_{G|X}$ is now given by $p(g|x) = \tr[M_g \rho_x]$ for a given measurement $\mathbb{M}=\{M_g\}$ and a given ensemble of states $\mathcal{E}=\{\rho_x, p(x)\}$. Operationally speaking, here the referee prepares the ensemble of states $\mathcal{E}$, sends one of these states to the gambler (say $\rho_x$), and then the gambler's goal is to try to identify the received state, by performing the measurement $\mathbb{M}=\{M_g\}$, and then using the outcome of this measurement to propose a betting strategy $b(x|g)$ \cite{DS2022}. In a similar manner, in a noisy quantum state betting (NQSB) game, the conditional PMF $p_{G|X}$ is given by $p(g|x) = \tr[M_g \mathcal{N}(\rho_x)]$, with the operational interpretation being same as for a QSB game, with the exception that the state the gambler receives is now affected by the (potentially noisy) channel $\mathcal{N}$ \cite{DS2022}. With these considerations in place, \cref{r:r7} gets translated as follows.

\begin{result} 
    \label{r:r11}
	Consider a QSB game defined by the pair $(o^{\sgn(q)c}_X, \mathcal{E})$ with constant odds as $o^{\sgn(q)c}(x) \coloneqq \sgn(q)\,C$, $C>0$, $\forall x$, and an ensemble of states $\mathcal{E} = \{\rho_x, p(x)\}$. Consider a first rational agent (gambler) represented by the isoelastic utility function $u_{1/q}$. We now compare the performance of this gambler when the gambler uses a fixed measurement $\mathbb{M}=\{M_g\}$, in comparison to being allowed to implement any possible uninformative measurement $\mathbb{N}\in{\rm UI}$. We remember here that the gamblers are interested in maximising the isoelastic certainty equivalent (ICE), and so in each case the gambler is allowed to play the betting task with the optimal betting strategy. Finally, consider a second rational agent represented by an isoelastic utility function $u_r$, who assesses the advantage (wealth ratio) achieved by the first $1/q$-agent. This latter assessment is given by the ID mutual information measure as:
	\begin{align}
    	&I^{\rm ID}_{q,r}
    	(X;G)_{\mathcal{E},\mathbb{M}}
    	\label{eq:result1}
        \\
        \nonumber
        &=
            \sgn(q)\,
            u_r
        	\left[
        	\frac{
        		\displaystyle
        		\max_{
        			b_{X|G}
        		}
        		\,
        		w^{\rm ICE}_{1/q}
        		\left(
        		b_{X|G}
        		,
        		\mathbb{M}
        		,
        		o_X^{\sgn(q)c}
        		,
        		\mathcal{E}
        		\right)
        	}{
        		\displaystyle
        		\max_{\mathbb{N}\in {\rm UI}}
        		\max_{
        			b_{X|G}
        		}
        		\,
        		w^{\rm ICE}_{1/q}
        		\left(
        		b_{X|G}
        		,
        		\mathbb{N}
        		,
        		o_X^{\sgn(q)c}
        		,
        		\mathcal{E}
        		\right)
        	}
        	\right]
        	.
    \end{align}
\end{result}
This result follows from \cref{r:r7} and from the observation that \emph{uninformative} measurements do not allow for the extraction of valuable information contained in the received state since $p(g|x)=\tr[M_g\rho_x]=q(g)\tr[\mathds{1}\rho_x]=q(g)$, and therefore we can check that:
\begin{align}
    &\max_{\mathbb{N}\in {\rm UI}}
	\max_{
		b_{X|G}
	}
	\,
	w^{\rm ICE}_{1/q}
	\left(
	b_{X|G}
	,
	\mathbb{N}
	,
	o_X^{\sgn(q)c}
	,
	\mathcal{E}
	\right)
	\\
	&=
	\max_{b_{X}}
	\,
	w^{\rm ICE}_{1/q}
	(
	b_X,
	o_X^{\sgn(q)c},
	p_X
	)
	.
	\nonumber
\end{align}
We now address the noisy quantum state betting (NQSB) games and the QRT of non-constant channels.

\begin{result}
    \label{r:r12}
    Consider a NQSB game defined by the tuple $(o^{\sgn(q)c}_X, \mathcal{E}, \mathcal{N})$ with constant odds as $o^{\sgn(q)c}(x) \coloneqq \sgn(q)\, C$, $C>0$, $\forall x$, an ensemble of states $\mathcal{E} = \{\rho_x,p(x)\}$, and a quantum channel $\mathcal{N}$. Consider a first rational agent (gambler) represented by the isoelastic utility function $u_{1/q}$. We now compare the performance of this gambler having access to the quantum channel $\mathcal{N}$, in comparison to being allowed to implement any possible constant channel $\mathcal{N}' \in \mathcal{N}$. In each case the gambler is allowed to play the betting task with the optimal betting strategy, and to optimise over all possible measurements. We also remember here that the gamblers are interested in maximising the isoelastic certainty equivalent (ICE). Finally, consider a rational agent represented by a utility function $u_r$, who assesses the performance of the advantage achieved by the $1/q$-agent. This latter assessment is given by the ID mutual information measure as:
	{\small
	\begin{align}
    	&I^{\rm ID}_{q,r}
    	(X;G)_{\mathcal{E},\mathcal{N}}
    	=
    	\label{eq:result5}
        \\
        \nonumber
        &
            \sgn(q)\,
            u_r
        	\left[
        	\frac{
        		\displaystyle
        		\max_{\mathbb{M}}
        		\max_{
        			b_{X|G}
        		}
        		\,
        		w^{\rm NQSB}_{1/q}
        		\left(
        		b_{X|G}
        		,
        		\mathbb{M}
        		,
        		o_X^{\sgn(q)c}
        		,
        		\mathcal{E}
        		,
        		\mathcal{N}
        		\right)
        	}{
        		\displaystyle
        		\max_{\substack{
        		\mathcal{N'}\in
        		\mathcal{C}
        		}}
        		\max_{\mathbb{N}}
        		\max_{
        			b_{X|G}
        		}
        		\,
        		w^{\rm NQSB}_{1/q}
        		\left(
        		b_{X|G}
        		,
        		\mathbb{N}
        		,
        		o_X^{\sgn(q)c}
        		,
        		\mathcal{E}
        		,
        		\mathcal{N}'
        		\right)
        	}
        	\right]
        	.
    \end{align}}
\end{result}
This result follows from \cref{r:r7} and from the observation that \emph{constant} channels are not able to extract information contained in the received state since $p(g|x)=\tr[M_g\mathcal{N}(\rho_x)]=\tr[M_g\sigma_\mathcal{N}]=q(g)$, and then we can check that:
\begin{align}
    &
    \max_{\substack{
	\mathcal{N'}\in
	\mathcal{C}
	}}
	\max_{\mathbb{N}}
	\max_{
		b_{X|G}
	}
	\,
	w^{\rm NQSB}_{1/q}
	\left(
	b_{X|G}
	,
	\mathbb{N}
	,
	o_X^{\sgn(q)c}
	,
	\mathcal{E}
	,
	\mathcal{N}'
	\right)
	\\
	&=
	\max_{b_{X}}
	\,
	w^{\rm ICE}_{1/q}
	(
	b_X,
	o_X^{\sgn(q)c},
	p_X
	)
	.
	\nonumber
\end{align}

\section{Conclusions}
\label{s:conclusions}

In this work we establish fundamental connections between utilities of wealth and information theory. Specifically, we derive results for general utility theories of \emph{wealth} as well as for utility theories of \emph{wealth ratios}.

First, regarding utility theories of \emph{wealth}, we introduce new operational tasks in the form of betting tasks in which \emph{both} gambler and bookmaker have access to side information, or betting tasks with \emph{double} side information for short. This, as an extension of works in the literature where it is usually assumed that \emph{only} the gambler has access to side information. In particular, we show the seemingly counterintuitive fact that, betting tasks with bookmaker's side information can be \emph{more} advantageous (for the gambler) than betting tasks \emph{without} any side information at all. This apparently paradoxical situation can be explained by remembering that whilst in this scenario the bookmaker is indeed the only one having access to side information, the bookmaker still needs to make the odds function public (as per the rules of the betting task) and therefore, the gambler can still implicitly have access to the side information in question. Notwithstanding this however, we moreover prove that this scenario \emph{cannot} be more advantageous (for the gambler) than receiving the side information directly. We prove this by linking the scenarios of bookmaker's side information and gambler's side information to conditional R\'enyi divergences, for which we can derive data processing inequalities and consequently, being able to explicitly compare the scenarios in question. We do this by introducing new conditional R\'enyi divergences that characterise such betting tasks, and by deriving some of their properties. This then provides operational interpretations to these conditional R\'enyi divergences as well. Furthermore, we extend some of these results, from the theory of expected utility, to \emph{prospect theory}, the latter being a theory in which decision-making agents are allowed to behave irrationally (with respect to their assessment of probabilities), albeit in a systematic manner. We prove that a specific R\'enyi conditional entropy quantifies such deviation from rationality.

Second, regarding utility theories of \emph{wealth ratios}, we provide an operational interpretation of the generalised $(q,r)$ mutual information measure \`a la Sharma-Mittal recently introduced by Ili\'c and Djordjevi\'c \cite{ID1}. Explicitly, it quantifies the utility of an $r$-agent on a wealth ratio that compares the performance achieved by a $1/q$-agent in two betting scenarios: i) using the best betting strategy \emph{with} side information, and ii) using the best betting strategy \emph{without} side information. In other words, it quantifies the utility of an $r$-agent on the advantage achieved (in betting games) by a $1/q$-agent that has access to side information. This interpretation comes in the form a correspondence that holds for general PMFs $p_{XG}$, as well as general orders $(q,r) \in \mathds{\overline R}\times \mathds{R}$. In particular, for orders $(q,q)$, $(q,1)$, $(r,1)$, the Ili\'c-Djordjevi\'c mutual information measure recovers the measures of Tsallis, Arimoto, and Gauss, respectively, and therefore, our result provides an operational interpretation for these quantities as well. This result also explicitly provides an operational meaning for the parameters $(q,r) \in \mathds{\overline R}\times \mathds{R}$, in terms of the \emph{risk-aversion} of rational agents playing betting tasks. In particular, whilst the $q$ parameter relates to the risk-aversion of a first rational agent in an \emph{inversely proportional} manner as $R_1=1/q$, the parameter $r$ on the other hand is found to be instead \emph{directly proportional} to the risk-aversion of a second rational agent as $R_2=r$. Moreover, we show that the Ili\'c-Djordjevi\'c conditional entropy satisfies a type of generalised chain rule which, as a particular case, recovers that of the Arimoto-R\'enyi conditional entropy. 

Finally, we address the implication of the results on wealth ratios on the quantum resource theories of measurement informativeness and non-constant channels. In the context of the QRT of measurement informativeness, and quantum state betting games, the Ili\'c-Djordjevi\'c (ID) mutual information measure quantifies the $r$-agent utility on the advantage achieved by a $1/q$-agent having access to the resource of informativeness, against all possible uninformative measurements. In a similar manner, for the operational tasks of noisy quantum state betting, the ID mutual information measure quantifies an advantage, but now in the context of a QRT of non-constant channels.

\section*{Acknowledgements}

We thank Francesco Buscemi and Valerio Scarani for insightful discussions. A.F.D. acknowledges support from the International Research Unit of Quantum Information, Kyoto University, the Center for Gravitational Physics and Quantum Information (CGPQI), and COLCIENCIAS 756-2016. P.S. acknowledges support from a Royal Society URF (NFQI). P.S. is a CIFAR Azrieli Global Scholar in the Quantum Information Science Programme. 

\appendix
\section{Additional details about some information-theoretic quantities}
\label{a:relationships}

\begin{definition} (Conditional-R\'enyi divergences \cite{thesis_CP}) 
	The Sibson conditional-R\'enyi divergence (S-CR-divergence) \cite{sibson}, the Csisz\'ar conditional-R\'enyi divergence (C-CR-divergence) \cite{csiszar}, and the Bleuler-Lapidoth-Pfister conditional-R\'enyi divergence (BLP-CR-divergence) \cite{BLP1} of order $\alpha \in \mathds{\overline R}$ of PMFs $p_{X|G}$, $q_{X|G}$, and $p_X$ are denoted as $D^{\rm V}_\alpha
	(
	p_{G|X}
	||
	q_{G|X}
	|
	p_X
	)$, with $\rm V\in\{S, C, BLP \}$. The orders $\alpha\in(-\infty,0)\cup(0,1)\cup(1,\infty)$ are defined as:
	\begin{align}
	&D^{\rm S}_\alpha
	(
	p_{G|X}
	||
	q_{G|X}
	|
	p_X
	)
	\label{eq:SCRD}
	\\
	&\coloneqq
	\frac{
	\sgn(\alpha)
	}{
	\alpha-1
	}
	\ln
	\left[
	\sum_x
	p(x)
	\sum_g
	p(g|x)^\alpha
	q(g|x)^{1-\alpha}
	\right]
	,
	\nonumber
	\\
	&D^{\rm C}_\alpha
	(
	p_{G|X}
	||
	q_{G|X}
	|
	p_X
	)
	\label{eq:CCRD}
	\\
	&\coloneqq
	\frac{
	\sgn(\alpha)
	}{
	\alpha-1
	}
	\sum_x
	p(x)
	\ln
	\left[
	\sum_g
	p(g|x)^\alpha
	q(g|x)^{1-\alpha}
	\right]
	,
	\nonumber
\end{align}
\begin{align}
	&D^{\rm BLP}_\alpha
	(
	p_{X|G}
	||
	q_{X|G}
	|
	p_G
	)
	\label{eq:BLPCRD2}
	\\
	&\coloneqq
	\frac{|\alpha|}{\alpha-1}
	\ln
	\left[
    	\sum_g
    	p(g)
    	\left(
        	\sum_x
        	p(x|g)^\alpha
        	q(x|g)^{1-\alpha}
    	\right)^{\frac{1}{\alpha}}	
	\right]
	.
	\nonumber
	\end{align}	
	The orders $\alpha \in\{1,0,\infty,-\infty\}$ are defined by their respective continuous extensions.
\end{definition}

\begin{definition} (Additional R\'enyi conditional entropies \cite{review_RCE}) R\'enyi conditional entropies of order $\alpha \in \mathds{\overline R}$ of a joint PMF $p_{XG}$ are denoted as $ H_{\alpha}^{j}(X|G)$, with $j\in\{1,4\}$, following the notation by Fehr and Berens \cite{review_RCE}. Arimoto's conditional entropy is denoted as $ H_{\alpha}^{\rm A}(X|G)$. The orders $\alpha \in (-\infty,0) \cup (0,1) \cup (1,\infty)$ are defined as:
\begin{align}
	H_{\alpha}^4
(X|G)
& \coloneqq
\frac{
	\sgn(\alpha)
}{
	1-\alpha
}
\ln
\left[  
\sum_g
p(g)
\sum_x
p(x|g)^\alpha
\right]
,
\\
H_{\alpha}^{1}
(X|G)
& \coloneqq
\sum_g
p(g)
\,
H_\alpha^{\rm R}
(X|G=g)
,
	\\
	H_{\alpha}^{\rm A}
	(X|G)
	& \coloneqq
	\frac{
		|\alpha|
	}{
		(1-\alpha)
	}
	\log
	\left[
	\sum_g
	\left(
	\sum_x
	p(x,g)^\alpha
	\right)^\frac{1}{\alpha}	
	\right]
	.
	\label{eq:ARCE}
\end{align}
	The orders $\alpha \in\{0,1,\infty,-\infty\}$ are defined by their respective continuous extensions.
\end{definition}

\begin{remark} (\cite{BLP1, thesis_CP})
    Relationship between CR-divergences and the R\'enyi divergence. For any conditional PMFs $p_{G|X}$, $q_{G|X}$, and any PMF $p_X$ we have:
    {\small\begin{align}
	&D^{\rm S}_\alpha
	(
	p_{G|X}
	||
	q_{G|X}
	|
	p_X
	)
    =
    D_\alpha(p_{G|X}p_X||q_{G|X}p_X)
    ,
	\label{eq:SR}
	\\
	&D^{\rm C}_\alpha
	(
	p_{G|X}
	||
	q_{G|X}
	|
	p_X
	)
	=
	\sum_x
	p(x)
	D_\alpha(p_{G|X=x}||q_{G|X=x})
	,
	\\
	&D^{\rm BLP}_\alpha
	(
	p_{G|X}
	||
	q_{G|X}
	|
	p_X
	)
	\nonumber
	\\
	&
	=
	\frac{
	\alpha
	}{
	\alpha-1
	}
	\log
	\left[
	\sum_x
	p(x)
	2^{
	\frac{\alpha-1}{\alpha}
	D_\alpha(p_{G|X=x}||q_{G|X=x})
	}
	\right]
	.
	\label{eq:CR}
	\end{align}}
\end{remark}

\begin{figure}[h!]
    \centering
    \includegraphics[scale=0.58]{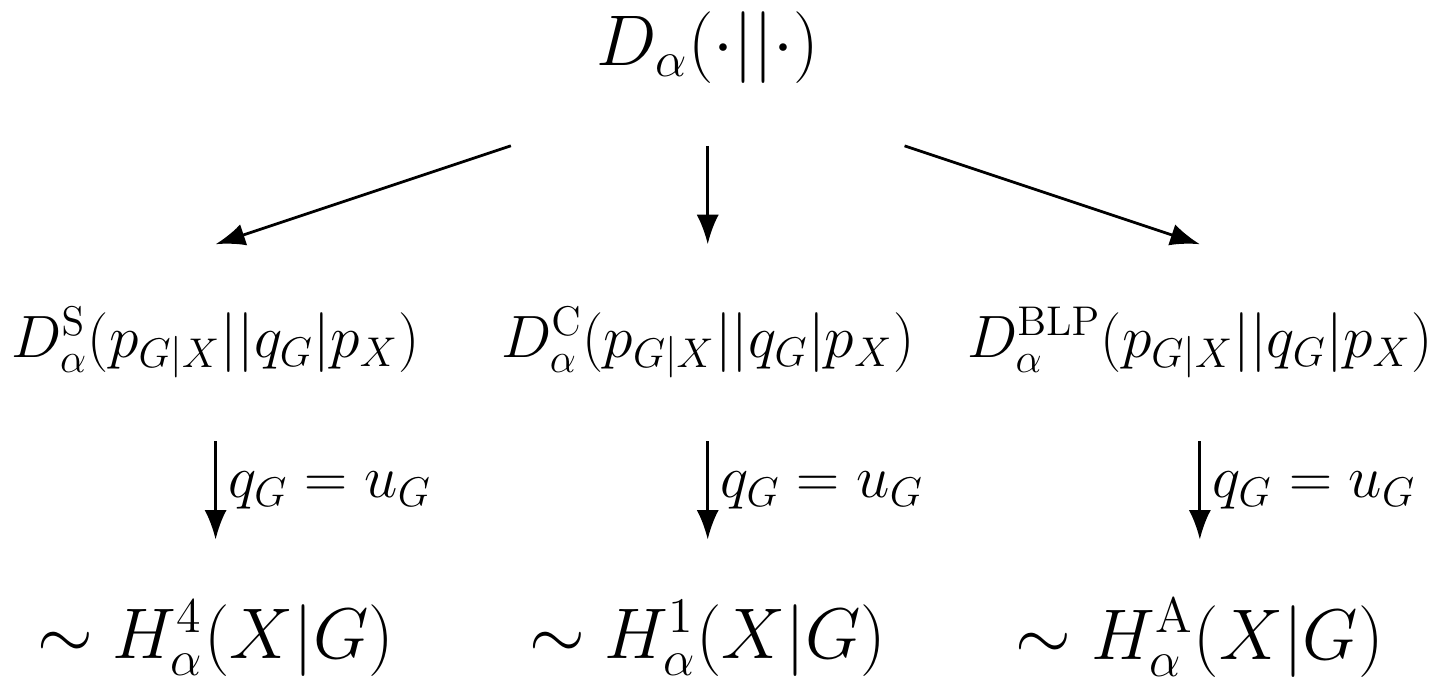}
    \vspace{-0.3cm}
    \caption{Hierarchical relationship between the R\'enyi divergence $D_\alpha(\cdot||\cdot)$, conditional R\'enyi divergences $D^{\rm V}_\alpha(\cdot||\cdot|\cdot)$, and conditional R\'enyi entropies $H^{\rm j}_\alpha(X|G)$. $\rm V\in \{S, C, BLP\}$ is a label specifying the measures of Sibson \cite{sibson}, Csisz\'ar \cite{csiszar}, and Bleuler-Lapidoth-Pfister \cite{BLP1}.}
    \label{fig:CR_divergences2}
\end{figure}

\begin{remark}
    (Relationship between CR-divergences and conditional entropies)
    \begin{align}
        D^{\rm S}_\alpha
    	(
    	p_{X|G}
    	||
    	u_{X}
    	|
    	p_G
    	)
    	&=
    	\sgn(\alpha)
    	\ln K
    	-
    	H_\alpha^{4}
    	(X|G)
    	,
    	\\
    	D^{\rm C}_\alpha
    	(
    	p_{X|G}
    	||
    	u_{X}
    	|
    	p_G
    	)
    	&=
    	\sgn(\alpha)
    	\ln K
    	-
    	H_\alpha^{1}
    	(X|G)
    	,
    	\\
    	D^{\rm BLP}_\alpha
    	(
    	p_{X|G}
    	||
    	u_{X}
    	|
    	p_G
    	)
    	&=
    	\sgn(\alpha)
    	\ln K
    	-
    	H_\alpha^{\rm A}
    	(X|G)
    	.
    \end{align}
    These identities can be seen as the R\'enyi conditional counterpart of the R\'enyi unconditional relationship:
    \begin{align}
        D_\alpha
    	(
    	p_{X}
    	||
    	u_{X}
    	)
    	&=
    	\sgn(\alpha)
    	\ln K
    	-
    	H_\alpha^{\rm R}
    	(X)
    	.
    \end{align}
\end{remark}
We address these two remarks in \autoref{fig:CR_divergences2}.

\section{Some mathematical preliminaries}
\label{a:r0}

In this appendix we address some preliminary mathematical tools to prove the results in the main text. We address $r$-deformed logarithms, rewrite some of the quantities of interest in terms of $r$-deformed logarithms, and address the operations of pseudo addition and pseudo substraction. Consider the r-deformed natural logarithm (r-logarithm) as:
    \begin{align}
        \ln_r
        (x)
        &\coloneqq
        \begin{cases}
        \ln x, 
        & 
        r=1,
        \\
        \frac{
        x^{1-r}
        -
        1
        }{
        1-r
        }
        ,
        & 
        r\neq 1,
    \end{cases}
    \end{align}
    for $x\geq 0$ and with the deformation parameter $r \in \mathds{R}$. We have the pseudo-additivity identity:
\begin{align}
    \ln_r
    (xy)
    =
    \ln_r
    (x)
    +
    \ln_r
    (y)
    +
    (1-r)
    \ln_r
    (x)
    \ln_r
    (y)
    .
\end{align}
It is also convenient to have:
    \begin{align}
        \ln_r
        \left(
        \frac{x}{y}
        \right)
        =
        \left[
        \ln_r(x)
        -
        \ln_r(y)
        \right]
        y^{1-r}
        .
    \end{align}
In particular we have:
\begin{align}
    \ln_r
    \left(
    \frac{1}{y}
    \right)
    =
    -
    \ln_r(y)
    \,
    y^{1-r}
    .
\end{align}
For the q-exponential we have:
    \begin{align}
        e_q^{x}e_q^{y}
        &=
        e_q^{x+y+(1-q)xy}
        ,
        \\
        e_q^{x}
        &\coloneqq
        \left(
        1+(1-q)x
        \right)^{\frac{1}{1-q}}
    \end{align}
These functions are important because many of the quantities considered in this work can be written in terms of $r$-deformed logarithms. For instance, the isoelastic utility function can be written as:
\begin{align}
    u_R(w)
    =
    \sgn(w)
    \ln_R(|w|)
    .
    \label{eq:isoln}
\end{align}  
It is also going to be useful to introduce the function:
\begin{align}
    \eta_r(x)
    \coloneqq
    \begin{cases}
        x, 
        & 
        r=1,
        \\
        \frac{
        e^{(1-r)x}
        -
        1
        }{
        1-r
        }
        ,
        & 
        r\neq 1.
    \end{cases}
\end{align}
This function in particular would update the standard natural logarithm into the $r$-deformed natural logarithm:
\begin{align}
    \eta_r
    (\ln x)
    =
    \ln_r(x)
    .
\end{align}
This function also allows for the manipulation of the pseudo-addition and pseudo-substraction introduced in the main text as:
\begin{align}
    x\oplus_r y
    &\coloneqq
    x+y+(1-r)xy,
    \\
    &=
    \eta_r
    \left(
        \eta_r^{-1}
        (x)
        +
        \eta_r^{-1}
        (y)
    \right)
    .
    \label{eq:pseudo_plus}
\end{align}
Similarly, the inverse of this operation, the pseudo-substraction, can be written as:
\begin{align}
    x\ominus_r y
    =
    \eta_r
    \left(
        \eta_r^{-1}
        (x)
        -
        \eta_r^{-1}
        (y)
    \right)
    .
    \label{eq:pseudo_minus}
\end{align}
Having introduced these functions, we can check that the Sharma-Mittal entropic quantities addressed in the main text adopt the following convenient functional forms:
\begin{align}
    H^{\rm SM}_{q,r}
	(X)
	&=
        \sgn(q)\,
	\eta_r
	\left(
            \sgn(q)\,
	    H^{\rm R}_q
	    (X)
	\right)
	,
	\label{eq:SM1}
	\\
	H^{\rm ID}_{q,r}
	(X|G)
	&=
        \sgn(q)\,
	\eta_r
	\left(
            \sgn(q)\,
	    H^{\rm A}_q
	    (X|G)
	\right)
	.
	\label{eq:IDCE1}
\end{align}
It is also useful to write:
\begin{align}
    H^{\rm R}_{q}
    (X)
    &=
        \sgn(q)\,
    \ln
    \left[
    p_q
    \left(
    X 
    \right)^{-1}
    \right]
    ,
    \label{eq:SM2}
    \\
    H^{\rm A}_{q}
    (X|G)
    &=
    \sgn(q)\,
    \ln
    \left[
    p_q
    \left(
    X|G
    \right)^{-1}
    \right]
    ,
    \label{eq:IDCE2}
\end{align}
with the quantities:
\begin{align}
	p_q
	\left(
	X 
	\right)
	& \coloneqq
	\left(
	\sum_x p(x)^q
	\right)^{\frac{1}{(q-1)}}
	\hspace{-0.6cm},
	\label{eq:rp}
	\\
	p_q
	\left(
	X|G
	\right)
	& \coloneqq
	\left(
	\sum_g
	\left(
	\sum_x
	p(x,g)^q
	\right)^\frac{1}{q}	
	\right)^\frac{q}{(q-1)}
	\hspace{-0.6cm}.
\end{align}
Finally, we will also need some useful inequalities. Jensen's inequality:
\begin{align}
    f\left(
        \mathbb{E}_{p_X}
        (X)
    \right)
    &\leq
    \mathbb{E}_{p_X}
    (
    f(X)
    )
    ,
    \hspace{1cm}
    \text{$f$ convex}
    .
\end{align}
H\"olders inequality:
\begin{align}
        \mathbb{E}
        (
        |XY|
        )
        &\leq
        \mathbb{E}
        (
        |X|^p
        )^\frac{1}{p}
        \mathbb{E}
        (
        |Y|^q
        )^\frac{1}{q}
        ,
\end{align}
for 
$p,q 
\in 
[1,\infty]
$
, 
$
\frac{1}{p}
+
\frac{1}{q}
=
1
$. Minkowski inequality:
{\small\begin{align}
    \left(
        \sum_x
        \left(
            \sum_y
            a(x,y)
        \right)^p 
        \,
    \right)^\frac{1}{p}
    \hspace{-0.3cm}
    &\leq
    \sum_y
    \left(
        \sum_x
        a(x,y)^p
    \right)^\frac{1}{p}
    \hspace{-0.3cm}
    ,
    \hspace{0.3cm}
    p>1,
    \\
    \left(
        \sum_x
        \left(
            \sum_y
            a(x,y)
        \right)^p 
        \,
    \right)^\frac{1}{p}
    \hspace{-0.3cm}
    &\geq
    \sum_y
    \left(
        \sum_x
        a(x,y)^p
    \right)^\frac{1}{p}
    \hspace{-0.3cm}
    ,
    \hspace{0.3cm}
    p<1,
\end{align}}
for non-negative values $\{a(x,y)\}$.

\section{Proof of \cref{r:r1}}
\label{a:r1}

\begin{proof}
Considering the PMFs:
    \begin{align}
        p(x)
        \coloneqq
        \sum_y
        p(x|y)
        p(y)
        ,
        \hspace{0.5cm}
        q(x)
        \coloneqq
        \sum_y
        q(x|y)
        p(y)
        .
    \end{align}
Consider now the R\'enyi divergence:
    {\small\begin{align}
        &
        e^{
        \sgn(\alpha)
        (\alpha-1)
        D_\alpha
    	(
    	p_{X}
    	||
    	q_{X}
    	)
        }
    	\\
    	&
            \overset{1}{=}
    	\sum_x
            	p(x)^\alpha\,
            	q(x)^{1-\alpha}
    	,
    	\\
    	&
            \overset{2}{=}
    	\sum_x
    	    \left(
    	    \sum_{y}
    	        p(y)
            	p(x|y)
                \right)^\alpha
            	\left(
            	    \sum_y
                    p(y)
                    q(x|y)
            	\right)^{1-\alpha}
    	\hspace{-0.3cm},
            \\
    	&
            \overset{3}{=}
    	\sum_x
    	    \left(
    	    \sum_{y}
    	        p(y)
            	(p(x|y)^\alpha)^\frac{1}{\alpha}
                \hspace{-0.1cm}
                \right)^{\hspace{-0.1cm}\alpha}
                \hspace{-0.2cm}
            	\left(
            	    \sum_y
                    p(y)
                    (q(x|y)^{1-\alpha})^\frac{1}{1-\alpha}
                    \hspace{-0.1cm}
            	\right)^{1-\alpha}
\hspace{-0.5cm},\\
    	&\overset{\rm (H)}{\geq}
    	\sum_x
    	    \sum_{y}
    	        p(y)\,
            	p(x|y)^\alpha
            	q(x|y)^{1-\alpha}
    	\hspace{-0.3cm},
            \\
    	&\overset{5}{=}
    	\sum_x
    	    \sum_{y}
            p(y)\,
            \left(    
            	p(x|y)
            	q(x|y)^\frac{1-\alpha}{\alpha}
             \right)^\alpha
    	,
            \\
    	&\overset{\rm (J)}{\geq}
    	\sum_x
            \left(
    	    \sum_{y}
    	        p(y)\,
            	p(x|y)
            	q(x|y)^\frac{1-\alpha}{\alpha}
             \right)^\alpha
    	,
        \\
        &\overset{7}{=}
        e^{
        \sgn(\alpha)
        (\alpha-1)
        D_\alpha^{\rm n1}
    	(
    	p_{X|Y}
    	||
    	q_{X|Y}
    	|
    	p_{Y}
    	)
        }
        .
\end{align}}
    The first line is the definition of the R\'enyi divergence. In the second line we replace the PMFs $p_X$ and $q_X$. In the third line we reorganise the expression so to use H\"older's inequality. In the fourth line we use H\"older's inequality with $p=1/\alpha$ and $q=1/(1-\alpha)$. In the fifth line we reorganise so to use Jensen's inequality. In the sixth line we use Jensen's inequality with the function $f(\cdot)=(\cdot)^\alpha$ which is convex on $\mathds{R}_{\geq 0}$. In the seventh line we identify the n1-CR-divergence of the conditional PMFs $p_{X|Y}$, $q_{X|Y}$, and PMF $p_Y$. The factor $\sgn(\alpha)(\alpha-1)$ is negative for $\alpha \in (0,1)$, so this reverses the inequality and get:
    \begin{align}
        D_\alpha
    	(
    	p_{X}
    	||
    	q_{X}
    	)
        \leq
        D_\alpha^{\rm n1}
    	(
    	p_{X|Y}
    	||
    	q_{X|Y}
    	|
    	p_{Y}
    	)
    .
    \end{align}
    This completes the proof.
\end{proof}

\section{Proof of \cref{r:r2}}
\label{a:r2}

\begin{proof}
    We start with the first new conditional R\'enyi divergence. Considering $\alpha>1$ we have:
    \begin{align}
        &
        \sgn(\alpha)
        (\alpha-1)
        D^{\rm n1}_\alpha
    	(
    	p_{X|Y}
    	||
    	q_{X|Y}
    	|
    	p_{Y}
    	)
    	\\
    	&
            \overset{1}{=}
    	\ln
    	\sum_x
    	    \left(
    	    \sum_y
    	        p(y)\,
            	p(x|y)\,
            	q(x|y)^\frac{1-\alpha}{\alpha}
        	\right)^\alpha
    	,
    	\\
    	&
            \overset{2}{=}
    	\alpha
    	\ln
    	\left[
    	\left(
    	\sum_x
    	    \left(
    	    \sum_y
    	        p(y)\,
            	p(x)\,
            	q(x|y)^\frac{1-\alpha}{\alpha}
        	\right)^\alpha
            \right)^\frac{1}{\alpha}
    	\right]
    	,
    	\\
    	&\overset{\rm (M)}{\leq}
    	\alpha
    	\ln
    	\sum_y
    	\left(
    	    \sum_x
    	    \left(
    	        p(y)\,
            	p(x|y)\,
            	q(x|y)^\frac{1-\alpha}{\alpha}
        	\right)^\alpha
        \right)^\frac{1}{\alpha}
    	,
    	\\
    	&
            \overset{4}{=}
    	\alpha
    	\ln
    	\sum_y
    	p(y)
    	\left(
    	    \sum_x
            	p(x|y)^\alpha\,
            	q(x|y)^{1-\alpha}
        \right)^\frac{1}{\alpha}
    	,
    	\\
    	&
            \overset{5}{=}
            \sgn(\alpha)
    	(\alpha-1)
            D^{\rm BLP}_\alpha
    	(
    	p_{X|Y}
    	||
    	q_{X|Y}
    	|
    	p_{Y}
    	)
    	.
	\end{align}
    In the first line we use the definition of the n1-CR-divergence. In the second line we multiply by $1=\alpha/\alpha$, leaving $1/\alpha$ inside the natural logarithm. In the third line we use Minkowski inequality with $p = \alpha$ (we are assuming here $\alpha>1$), and the coefficients $a(x,y) = p(y)\,p(x|y)\,q(x|y)^\frac{1-\alpha}{\alpha}$, and we also take into account that $\log (\cdot)$ is increasing. In the fourth line we reorganise. In the fifth line we identify the BLP-CR-divergence. This then proves the case for $\alpha>1$. We now analyse the additional cases.
    
    There are three factors to consider when analysing the direction of the inequality. First, whether $\alpha>1$ or $\alpha<1$ so to use Minkowski inequality. Second, whether $\alpha >0$ or $\alpha <0$ because this factor multiplies the whole expression. Third, the sign of the factor $\sgn(\alpha)(\alpha-1)$. Considering now the case $\alpha<1$ we have that the Minkowski inequality gets reversed, and depending on $\sgn(\alpha)$ we either keep the inequality or further reverse it again, and then a similar analysis with the factor $\sgn(\alpha)(\alpha-1)$. In summary, a careful inspection of the two remaining cases ($\alpha<0$, and $0<\alpha<1$) yield the inequality:
    \begin{align}
        D^{\rm n1}_\alpha
        &\leq
        D^{\rm BLP}_\alpha
        ,
        \hspace{0.3cm}
        \alpha
        \in
        (-\infty,0)
        \cup(0,1)
        \cup
        (1,\infty)
        .
    \end{align}
\end{proof}

\onecolumngrid
\section{Proof of \cref{r:r3}}
\label{a:r3}

\begin{proof}
    From the definition of PMF $h_X$ we have:
    \begin{align}
        h(x)^{R}
	\coloneqq
	\frac{
	    \sum_y
		p(xy)
		|o(x|y)|^{1-R}
	}
	{
	\left(
	    \sum_{x'}
			\left(
    		    \sum_y
    			p(x'y)
    			\,
    			|o(x'|y)|^{1-R}
			\right)^\frac{1}{R}
	\right)^{R}
	}.  
    \end{align}
     and so:
     \begin{align}
        \sum_y
        p(xy)\,
		|o(x|y)|^{1-R}
		=
		\left(
	    \sum_{x'}
			\left(
    		    \sum_y
    			p(x'y)
    			\,
    			|o(x'|y)|^{1-R}
			\right)^\frac{1}{R}
	    \right)^{R}
		h(x)^{R}
		.
     \end{align}
    Then, the natural logarithm of the CE can be written as:
     \begin{align}
         \ln
         \left[
         w^{\rm ICE}_{R}
         (b_{X}, |o_{X|Y}|, p_{XY})
         \right]
         &\overset{1}{=}
         \frac{1}{1-R}
         \ln
         \left[
         \sum_{xy}
            p(x,y)\,
            |o(x|y)|^{1-R}
            b(x)^{1-R}
        \right]
        ,
        \\
         &\overset{2}{=}
         \frac{1}{1-R}
         \ln
         \left[
         \sum_{x}
            \left(
                \sum_y
                p(x,y)\,
                |o(x|y)|^{1-R}
            \right)
            b(x)^{1-R}
        \right]
        ,
        \\
        &\overset{3}{=}
        \frac{1}{1-R}
         \ln
         \left[
         \sum_{x}
        \left(
            \sum_{x'}
                \left(
                    \sum_y
                    p(x'y)
                    \,
                    |o(x'|y)|^{1-R}
                \right)^\frac{1}{R}
		\right)^{R}
		    h(x)^{R}
            b(x)^{1-R}
        \right]
        ,
        \\
        &\overset{4}{=}
        \frac{R}{1-R}
         \ln
         \left[
        \sum_{x'}
            \left(
                \sum_y
                p(x'y)
                \,
                |o(x'|y)|^{1-R}
            \right)^\frac{1}{R}
    \right] 
		+
		\frac{1}{1-R}
		\ln
		\left[
		\sum_{x}
    		h(x)^{R}
            b(x)^{1-R}
        \right]
        ,
		\\
        &\overset{5}{=}
        \frac{R}{1-R}
         \ln
         \left[
		    \sum_{x'}
    			\left(
        		    \sum_y
        			p(x'y)
        			\,
        			r(x'|y)^{-(1-R)}
			    \right)^\frac{1}{R}
		\right] 
		-
            \sgn(R)
		D_{R}
		(
		h^{(R, o, p)}_{X}
		||
		b_{X}
		)
		,
		\\
        &\overset{6}{=}
        \sgn(R)
        D_{1/R}^{\rm n1}
        (
        p_{X|Y}
        ||
        r^{(o)}_{X|Y}
        |
        p_{Y}
        )
        -
        \sgn(R)
        D_{R}
        (
        h^{(R, o, p)}_{X}
        ||
        b_{X}
        )
        .
\end{align}
In the first line we use the definition of the ICE. In the second line we group terms and sum over $y$. In the third line we replace the previous equality. In the fourth line we reorganise. In the fifth line we identify the R\'enyi divergence of order $R$. In the sixth and final line we replace the PMF $r_{X|Y}$ and identify the $\rm n1$-CR divergence. We can then multiply both sides of the equality by $\sgn(o)$ and this finishes the proof.
\end{proof}

\section{Proof of \cref{r:r4}}
\label{a:r4}

\begin{proof}
    From the first PMF we have:
    \begin{align}
        h(x|g)^{R}
	\coloneqq
	\frac{
	    \sum_y
		p(xy|g)
		o(x|y)^{1-R}
	}
	{
	\left(
	    \sum_{x'}
			\left(
    		    \sum_y
    			p(x'y|g)
    			\,
    			o(x'|y)^{1-R}
			\right)^\frac{1}{R}
	\right)^{R}
	}.  
    \end{align}
     and so:
     \begin{align}
        \sum_y
        p(xy|g)\,
		o(x|y)^{1-R}
		=
		\left(
	    \sum_{x'}
			\left(
    		    \sum_y
    			p(x'y|g)
    			\,
    			o(x'|y)^{1-R}
			\right)^\frac{1}{R}
	    \right)^{R}
		h(x|g)^{R}
		.
     \end{align}
     Multiplying by $p(g)$ we get:
     \begin{align}
        p(g)
        \left(
            \sum_y
            p(xy|g)\,
    		o(x|y)^{1-R}
		\right)
		=
		p(g)
		\left(
	    \sum_{x'}
			\left(
    		    \sum_y
    			p(x'y|g)
    			\,
    			o(x'|y)^{1-R}
			\right)^\frac{1}{R}
	    \right)^{R}
		h(x|g)^{R}
		.
     \end{align}
     From the second PMF we get:
     \begin{align}
         h
		(g)
		&\coloneqq 
		\frac{
		    p(g)
		    \left[
		    \sum_{x}
    			\left(
        		    \sum_y
        			p(xy|g)
        			\,
        			o(x|y)^{1-R}
			    \right)^\frac{1}{R}
			\right]^{R}
		}
		{
			\sum_{g'}
			p(g')
		    \left[
		    \sum_{x}
    			\left(
        		    \sum_y
        			p(xy|g')
        			\,
        			o(x|y)^{1-R}
			    \right)^\frac{1}{R}
			\right]^{R}
		}.
     \end{align}
     and so:
     \begin{align}
         	h(g)
         	\left(
             	\sum_{g'}
    			p(g')
    		    \left[
    		    \sum_{x}
        			\left(
            		    \sum_y
            			p(xy|g')
            			\,
            			o(x|y)^{1-R}
    			    \right)^\frac{1}{R}
    			\right]^{R}
			\right)
		    &=
		    p(g)
		    \left[
		    \sum_{x}
    			\left(
        		    \sum_y
        			p(xy|g)
        			\,
        			o(x|y)^{1-R}
			    \right)^\frac{1}{R}
			\right]^{R}
			.
     \end{align}
    The natural logarithm of the CE can be written as:
     {\small
     \begin{align}
         &
         \ln
         \left[
         w^{\rm ICE}_{R}
         (b_{X|G}, |o_{X|Y}|, p_{XGY})
         \right]
         \nonumber
         \\
         &\overset{1}{=}
         \frac{1}{1-R}
         \ln
         \left[
         \sum_{xgy}
            p(x,g,y)\,
            |o(x|y)|^{1-R}
            b(x|g)^{1-R}
        \right]
        ,
        \\
         &\overset{2}{=}
         \frac{1}{1-R}
         \ln
         \left[
         \sum_{xg}
            p(g)
            \left(
                \sum_y
                p(x,y|g)\,
                |o(x|y)|^{1-R}
            \right)
            b(x|g)^{1-R}
        \right]
        ,
        \\
        &\overset{3}{=}
        \frac{1}{1-R}
         \ln
         \left[
         \sum_{xg}
         h(g)
         	\left(
             	\sum_{g'}
    			p(g')
    		    \left[
    		    \sum_{x}
        			\left(
            		    \sum_y
            			p(xy|g')
            			\,
            			|o(x|y)|^{1-R}
    			    \right)^\frac{1}{R}
    			\right]^{R}
			\right)
		    h(x|g)^{R}
            b(x|g)^{1-R}
        \right]
        ,
        \\
        &\overset{4}{=}
        \frac{1}{1-R}
         \ln
         \left[
         	\sum_{g'}
    			p(g')
    		    \left[
    		    \sum_{x}
        			\left(
            		    \sum_y
            			p(xy|g')
            			\,
            			|o(x|y)|^{1-R}
    			    \right)^\frac{1}{R}
    			\right]^{R}
		\right] 
		+
		\frac{1}{1-R}
		\ln
		\left[
		\sum_{xg}
    		h(x|g)^{R}
    		h(g)^{R}
            b(x|g)^{1-R}
            h(g)^{1-R}
        \right]
        ,
		\\
        &\overset{5}{=}
        \frac{1}{1-R}
         \ln
         \left[
		    \sum_{g'}
    			p(g')
    		    \left[
    		    \sum_{x}
        			\left(
            		    \sum_y
            			p(xy|g')
            			\,
            			|o(x|y)|^{1-R}
    			    \right)^\frac{1}{R}
    			\right]^{R}
		\right]
		-
            \sgn(R)
		D_{R}
		(
		h^{(R, o, p)}_{X|G}
		\,
		h^{(R, o, p)}_{G}
		||
		b_{X|G}
		\,
		h^{(R, o, p)}_{G}
		)
		,
		\\
        &\overset{6}{=}
        \frac{1}{1-R}
         \ln
         \left[
		    \sum_{g'}
    			p(g')
    		    \left[
    		    \sum_{x}
        			\left(
            		    \sum_y
            			p(xy|g')
            			\,
            			r(x|y)^{-(1-R)}
    			    \right)^\frac{1}{R}
    			\right]^{R}
		\right]
		-
            \sgn(R)
		D_{R}
		(
		h^{(R, o, p)}_{X|G}
		\,
		h^{(R, o, p)}_{G}
		||
		b_{X|G}
		\,
		h^{(R, o, p)}_{G}
		)
		,
		\\
        &\overset{7}{=}
        \sgn(R)
        D_{1/R}^{\rm n2}
    	(
    	p_{X|GY}
    	||
    	r^{(o)}_{X|Y}
    	|
    	p_{G}
    	,
    	p_{Y|G}
    	)
    	-
            \sgn(R)
		D_{R}
		(
		h^{(R, o, p)}_{X|G}
		\,
		h^{(R, o, p)}_{G}
		||
		b_{X|G}
		\,
		h^{(R, o, p)}_{G}
		)
		.
\end{align}}
In the first line we use the definition of the ICE. In the second line we group terms and sum over $y$. In the third line we replace the previous equality. In the fourth line we reorganise. In the fifth line we identify the R\'enyi divergence of order $R$. In the sixth and final line we replace the PMF $r_{X|Y}$ and identify the $\rm n1$-CR divergence. We can then multiply both sides of the equality by $\sgn(o)$ and this finishes the proof.
\end{proof}

\section{Details of \cref{r:c4}}
\label{a:c4}

\begin{proof}
We have the following chain of equalities:
    \begin{align}
	&\sgn(R)
	\ln
	\left[
	\frac{
		\max_{b_{X}}
		w^{\rm ICE}_{R}
		(
		b_{X}
		,
		o_{X|Y}
		,
		p_{XY}
		)
	}{
		\displaystyle
		\max_{b_X}
		w^{\rm ICE}_{R}
		(
		b_X
		,
		o_X
		,
		p_X
		)
	}
	\right]
	\nonumber
	\\
	&\overset{1}{=}
	\sgn(R)
	\ln
	\left[
	\frac{
	    \sgn(R)
		\max_{b_{X}}
		\sgn(R)\,
		w^{\rm ICE}_{R}
		(
		b_{X}
		,
		|o_{X|Y}|
		,
		p_{XY}
		)
	}{
		\displaystyle
		\sgn(R)
		\max_{b_X}
		\sgn(R)\,
		w^{\rm ICE}_{R}
		(
		b_X
		,
		|o_X|
		,
		p_X
		)
	}
	\right]
	,
        \\
	&\overset{2}{=}
	\sgn(R)
	\ln
	\left[ 
        \sgn(R)
        \max_{b_X}
        \sgn(R)
		w^{\rm ICE}_{R}
		(
		b_{X}
		,
		|o_{X|Y}|
		,
		p_{XY}
		)
	\right]
	-
	\sgn(R)
	\ln
	\left[ 
        \sgn(R)
        \max_{b_X}
        \sgn(R)
		w^{\rm ICE}_{R}
		(
		b_X
		,
		|o_X|
		,
		p_X
		)
	\right]
	,
	\\
	&\overset{3}{=}
	\max_{b_{X}}
	\sgn(R)
	\ln
	\left[ 
		w^{\rm ICE}_{R}
		(
		b_{X}
		,
		|o_{X|Y}|
		,
		p_{XY}
		)
	\right]
	-
	\max_{b_X}
	\sgn(R)
	\ln
	\left[ 
		w^{\rm ICE}_{R}
		(
		b_X
		,
		|o_X|
		,
		p_X
		)
	\right]
	,
	\\
	&\overset{4}{=}
	\max_{b_{X}}
	U^{\rm ICE}_{R}
         (
         b_{X}
         ,
         o_{X|Y}
         , 
         p_{XY}
         )
    -
    \max_{b_{X}}
	U^{\rm ICE}_{R}
         (
         b_{X}
         ,
         o_{X}
         ,
         p_{X}
         )
    ,\\
    &\overset{5}{=}
    D_{1/R}^{
    \rm n1
    }
    (
    p_{X|Y}
    ||
    r^{(o)}_{X|Y}
    |
    p_Y
    )
    -
    D_{1/R}
    (
    p_X
    ||
    r_X^{(o)}
    )
    .
\end{align}
In the first line we multiply and divide by $\sgn(R)$ inside the natural logarithm, and take out $\sgn(o)=\sgn(R)$ from the ICE, leaving the ICE with positive odds. In the second line we split the natural logarithm. In the third line we take out the $\max_{b_X}$, which jumps through two $\sgn(R)$, and therefore keeps being a maximisation. In the third line we remember that $\sgn(o)=\sgn(R)$ and identify the functions $U_R^{\rm ICE}$. In the fourth line we use \cref{c:c1} and Corollaries 6 and 7 from \cite{DS2022}. This finishes the statement.
\end{proof}

\section{Details of \cref{r:c5}}
\label{a:c5}

\begin{proof}
We have the following chain of equalities:
    \begin{align}
	&\sgn(R)
	\ln
	\left[
	\frac{
		\max_{b_{X|G}}
		w^{\rm ICE}_{R}
		(
		b_{X|G}
		,
		o_{X}
		,
		p_{XG}
		)
	}{
		\displaystyle
		\max_{b_X}
		w^{\rm ICE}_{R}
		(
		b_X
		,
		o_{X|G}
		,
		p_{XG}
		)
	}
	\right]
	\nonumber
	\\
	&\overset{1}{=}
	\sgn(R)
	\ln
	\left[
	\frac{
	    \sgn(R)
		\max_{b_{X}}
		\sgn(R)\,
		w^{\rm ICE}_{R}
		(
		b_{X|G}
		,
		|o_{X}|
		,
		p_{XG}
		)
	}{
		\displaystyle
		\sgn(R)
		\max_{b_X}
		\sgn(R)\,
		w^{\rm ICE}_{R}
		(
		b_X
		,
		|o_{X|G}|
		,
		p_{XG}
		)
	}
	\right]
	,
        \\
	&\overset{2}{=}
	\sgn(R)
	\ln
	\left[ 
        \sgn(R)
	\max_{b_{X|G}}
	\sgn(R)
		w^{\rm ICE}_{R}
		(
		b_{X|G}
		,
		|o_{X}|
		,
		p_{XG}
		)
	\right]
	-
	\sgn(R)
	\ln
	\left[ 
        \sgn(R)
	\max_{b_{X|G}}
	\sgn(R)
		w^{\rm ICE}_{R}
		(
		b_X
		,
		|o_{X|G}|
		,
		p_{XG}
		)
	\right]
	,
	\\
	&\overset{3}{=}
	\max_{b_{X|G}}
	\sgn(R)
	\ln
	\left[ 
		w^{\rm ICE}_{R}
		(
		b_{X|G}
		,
		|o_{X}|
		,
		p_{XG}
		)
	\right]
	-
	\max_{b_X}
	\sgn(R)
	\ln
	\left[ 
		w^{\rm ICE}_{R}
		(
		b_X
		,
		|o_{X|G}|
		,
		p_{XG}
		)
	\right]
	,
	\\
	&\overset{4}{=}
	\max_{b_{X|G}}
	U^{\rm ICE}_{R}
         (
         b_{X|G}
         ,
         o_{X}
         , 
         p_{XG}
         )
    -
    \max_{b_{X}}
	U^{\rm ICE}_{R}
         (
         b_{X}
         ,
         o_{X|G}
         ,
         p_{XG}
         )
    ,\\
    &\overset{5}{=}
    D_{1/R}^{
    \rm BLP
    }
    (
    p_{X|G}
    ||
    r^{(o)}_{X|G}
    |
    p_G
    )
    -
    D_{1/R}^{
    \rm n1
    }
    (
    p_{X|G}
    ||
    r^{(o)}_{X|G}
    |
    p_G
    )
    .
\end{align}
In the first line we multiply and divide by $\sgn(R)$ inside the natural logarithm, and take out $\sgn(o)=\sgn(R)$ from the ICE, leaving the ICE with positive odds. In the second line we split the natural logarithm. In the third line we take out the $\max_{b_X}$, which jumps through two $\sgn(R)$, and therefore keeps being a maximisation. In the third line we remember that $\sgn(o)=\sgn(R)$ and identify the functions $U_R^{\rm ICE}$. In the fourth line we use \cref{c:c1} and Corollaries 6 and 7 from \cite{DS2022}. This finishes the statement.
\end{proof}

\vspace{-1cm}
\section{Proof of \cref{r:r5} (PT1)}
\label{a:r5}

\begin{proof}
    Consider the natural logarithm of the certainty equivalent within prospect theory:
     \begin{align}
         \ln
         \left[
         V^{\rm CE-PT}_{R, S}
         (
         b_{X}
         ,
         |o_X|
         ,
         p_{X}
         )
         \right]
         &\overset{1}{=}
         \frac{1}{1-R}
         \ln
         \left[
         \sum_{x}
            p(x)^S
            |o(x)|^{1-R}
            b(x)^{1-R}
        \right]
        ,
        \\
         &\overset{2}{=}
         \frac{1}{1-R}
         \ln
         \left[
         \left(
            \sum_{x'}
            p(x')^S
         \right)
         \sum_{xg}
            \frac{
            p(x)^S
            }
            {
            \sum_{x'}
            p(x')^S
            }
            |o(x)|^{1-R}
            b(x)^{1-R}
        \right]
        ,
        \\
         &\overset{3}{=}
         \frac{1}{1-R}
         \ln
         \left[
            \sum_{x'}
            p(x')^S
         \right]
         +
         \frac{1}{1-R}
         \ln
         \left[
         \sum_{x}
            q(x)\,
            |o(x)|^{1-R}
            b(x)^{1-R}
        \right]
        ,
        \\
         &\overset{4}{=}
         \sgn(S)
         \frac{1-S}{1-R}
         H_S(p_{X})
         +
         \ln
         \left[
         V^{\rm CE-EU}_{R}
         \left(
         b_{X}, |o_X|, q_{X}^{(S,p)}
         \right)
         \right]
        .
    \end{align}
    The first equality is the certainty equivalent within prospect theory. In the second inequality we multiply and divide by the coefficient $\sum_{x'} p(x')^{S}$. In the third equality we reorganise the expression and identify the escort PMF $q_{X}^{(S,p)}$. In the fourth equality we identify the R\'enyi entropy of order $S$ and the certainty equivalent within expected utility theory. The final step is to input the expression for the CE-EU in terms of the R\'enyi divergence, and this finishes the proof. 
\end{proof}

\vspace{-1cm}
\section{Proof of \cref{r:r6} (PT2)}
\label{a:r6}

\begin{proof}
    Consider the natural logarithm of the certainty equivalent within prospect theory:
     \begin{align}
         \ln
         \left[
         V^{\rm CE-PT}_{R, S}
         (b_{X|G}, |o_X|, p_{XG})
         \right]
         &\overset{1}{=}
         \frac{1}{1-R}
         \ln
         \left[
         \sum_{xg}
            p(x,g)^S
            |o(x)|^{1-R}
            b(x|g)^{1-R}
        \right]
        ,
        \\
         &\overset{2}{=}
         \frac{1}{1-R}
         \ln
         \left[
         \left(
            \sum_{x'g'}
            p(x',g')^S
         \right)
         \sum_{xg}
            \frac{
            p(x,g)^S
            }
            {
            \sum_{x'g'}
            p(x',g')^S
            }
            |o(x)|^{1-R}
            b(x|g)^{1-R}
        \right]
        ,
        \\
         &\overset{3}{=}
         \frac{1}{1-R}
         \ln
         \left[
            \sum_{x'g'}
            p(x',g')^S
         \right]
         +
         \frac{1}{1-R}
         \ln
         \left[
         \sum_{xg}
            q(x,g)\,
            |o(x)|^{1-R}
            b(x|g)^{1-R}
        \right]
        ,
        \\
         &\overset{4}{=}
         \sgn(S)
         \frac{1-S}{1-R}
         H_S(p_{XG})
         +
         \ln
         \left[
         V^{\rm CE-EU}_{R}
         \left(
         b_{X|G}, |o_X|, q_{XG}^{(S,p)}
         \right)
         \right]
        .
    \end{align}
    The first equality is the certainty equivalent within prospect theory. In the second inequality we multiply and divide by the coefficient $\sum_{x'g'} p(x',g')^{S}$. In the third equality we reorganise the expression and identify the escort PMF $q_{XG}^{(S)}$. In the fourth equality we identify the R\'enyi entropy of order $S$ and the certainty equivalent within expected utility theory. The final step is input the known expression for this latter CE in terms of the R\'enyi divergence and the BLP conditional R\'enyi divergence, and this finishes the proof. 
\end{proof}

\twocolumngrid
\section{Proof of \cref{r:r7}}
\label{a:r7}

We first need the following lemmas.
\begin{lemma}
\label{lemma4}
    (\cite{DS2022})
	 Consider a PMF $p_X$, the R\'enyi probability of order $q \in \mathds{\overline R}$ can be written as:
	\begin{align}
	\sgn(q)
	\,
    C
    \,
	p_q(X)
	=
    \max_{b_X}
    w^{\rm ICE}_{1/q}
    (
    b_{X}
    ,
    o_X^{\sgn(q)c}
    ,
    p_X)
    ,
    \label{eq:lemmaR}
    \end{align}
	with the maximisation over all possible betting strategies $b_X$, and the odds $o^{\sgn(q)c}(x) \coloneqq \sgn(q)\,C$, $C>0$, $\forall x$.
\end{lemma}

\begin{lemma}(\cite{DS2022}) \label{lemma5}
	The Arimoto-R\'enyi conditional entropy of order $q \in \mathds{\overline R}$ for a joint PMF $p_{XG}$ reads:
	{\small \begin{align}
	\sgn(q)
	\,
	C
	\,
	p_q(X|G)
	=
    \max_{b_{X|G}}
    w^{\rm ICE}_{1/q}
    (
    b_{X|G}
    ,
    o_X^{\sgn(q)c}
    ,
    p_{XG}
    )
    ,
    \label{eq:lemmaAR}
    \end{align}}
	with the maximisation over all possible betting strategies $b_{X|G}$, and the odds $o^{\sgn(q)c}(x) \coloneqq \sgn(q)\,C$, $C>0$, $\forall x$.
\end{lemma}

\begin{proof}(of \cref{r:r7})
    Consider the Ili\'c-Djordjevi\'c mutual information, we have the following chain of equalities:
	{\small
	\begin{align}
	&I_{q,r}^{\rm ID}
	(X;G)
	\nonumber
	\\
	&
	=
	H_{q,r}^{\rm SM}(X)
	\ominus_r
	H_{q,r}^{\rm ID}(X|G)
	,
	\\
	&=
	\sgn(q)
	[
	\eta_r
	\left(
        \sgn(q)
	H_{q}^{\rm R}(X)
	\right)
	\ominus_r
	\eta_r
	\left(
        \sgn(q)
	H_{q}^{\rm A}(X|G)
	\right)
	]
	,
	\\
	&=
	\sgn(q)
	[
	\eta_r
	\left(
    	\ln 
    	\left[ 
    	p_q^{-1}(X)
    	\right]
	\right)
	\ominus_r
	\eta_r
	\left(
    	\ln 
    	\left[ 
    	p_q^{-1}(X|G)
    	\right]
	\right)
	]
	.
    \end{align}
    }
The first line is the definition of the Ili\'c-Djordjevi\'c mutual information. In the second line we use \eqref{eq:SM1} and \eqref{eq:IDCE1}. In the third line we use \eqref{eq:SM2} and \eqref{eq:IDCE2}. We take now $a = \eta_r
	\left(
    	\ln 
    	\left[ 
    	p_q^{-1}(X)
    	\right]
	\right)$
and
$b = \eta_r
	\left(
    	\ln 
    	\left[ 
    	p_q^{-1}(X|G)
    	\right]
	\right)$
and apply the rule for $a\ominus_r b$ \eqref{eq:pseudo_minus} and get:
\begin{align}
    &I_{q,r}^{\rm ID}
	(X;G)
	\nonumber
	\\
	&
	\overset{1}{=}
	\sgn(q)
	[
	\eta_r
	\left(
    	\ln 
    	\left[ 
    	p_q^{-1}(X)
    	\right]
	-
    	\ln 
    	\left[ 
    	p_q^{-1}(X|G)
    	\right]
	\right)
	]
	,
	\\
	&
	\overset{2}{=}
	\sgn(q)\,
	\eta_r
	\left(
	\ln
	\left[
	\frac{
		p_{q}
		(
		X|G
		)
	}
	{
		p_{q}	
		(
		X 
		)
	}
	\right]
	\right)
	,
	\\
	&
	\overset{3}{=}
	\sgn(q)
	\ln_r
	\left[
	\frac{
		p_{q}
		(
		X|G
		)
	}
	{
		p_{q}	
		(
		X 
		)
	}
	\right]
	,
	\\
	&
	\overset{4}{=}
	\sgn(q)
	\ln_r
	\left[
	\frac{
		\sgn(q)\,C\,p_{q}
		(
		X|G
		)
	}
	{
		\sgn(q)
		\,C\,p_{q}	
		(
		X 
		)
	}
	\right]
	,\\
	&
	\overset{5}{=}
	\sgn(q)\,
	u_r
	\left[
	\frac{
		\displaystyle
		\max_{b_{X|G}}
		w^{\rm ICE}_{1/q}
		(
		b_{X|G}
		,
		o_X^{\sgn(q)}
		,
		p_{XG}
		)
	}{
		\displaystyle
		\max_{b_X}
		w^{\rm ICE}_{1/q}
		(
		b_X
		,
		o_X^{\sgn(q)}
		,
		p_X
		)
	}
	\right]
	.
	\end{align}
	The first line follows from the chain of identities in the previous paragraph. In the second line we reorganise using properties of $\ln$. In the third line we use that $\eta_r(\ln (x))=\ln_r(x)$. In the fourth line we multiply and divide by $\sgn(q)\,C$. In the fifth line we use \eqref{eq:lemmaR} and \eqref{eq:lemmaAR}. This completes the proof.
\end{proof}

\section{Proof of \cref{r:r8}}
\label{a:r8}

\begin{proof}
    The inequality we want to prove is equivalent to the following inequalities. Consider $q\geq 0$:
    {\small
    \begin{align}
        \sgn(q)
        H^{\rm ID}_{q,r}(X|G)
        &\overset{1}{\geq}
        \sgn(q)
        \left[
            H^{\rm SM}_{q,r}(XG)
            \ominus_r
            \ln_r
            \left(
            K
            \right)
        \right]
        ,
    \\
        H^{\rm ID}_{q,r}(X|G)
        &\overset{2}{\geq}
        H^{\rm SM}_{q,r}(XG)
        \ominus_r
        \ln_r
        \left(
        K
        \right)
        ,
        \\
        \ln_r
    	\left[
    	p_q
    	\left(
    	X|G
    	\right)^{-1}
    	\right]
        &\overset{3}{\geq}
        \eta_r
        \left(
        \ln
        	\left[
        	p_q
        	\left(
        	XG
        	\right)^{-1}
        	\right]
    	\right)
        \ominus_r
        \eta_r
        \left(
        \ln
        \left[
        K
        \right]
        \right)
        ,
    \end{align}}	
    The first line is the inequality we want to prove. In the second line we assume $q\geq 0$ and cancel $\sgn(q)$. In the third line we use \eqref{eq:IDCE1}. We now apply the rule for $a\ominus_r b$ \eqref{eq:pseudo_minus} and we get:
    \begin{align}
        \ln_r
    	\left[
    	p_q
    	\left(
    	X|G
    	\right)^{-1}
    	\right]
        &\overset{4}{\geq}
        \eta_r
        \left(
    	\ln
    	\left[
    	p_q
    	\left(
    	XG
    	\right)^{-1}
    	\right]
    	-
    	\ln
        \left[
        K
        \right]
        \right)
    	,
    	\\
    	&
    	\overset{5}{\geq}
    	\eta_r
    	\ln
    	\left[
    	p_q
    	\left(
    	XG
    	\right)^{-1}
    	K^{-1}
    	\right]
    	,
    	\\
    	&
    	\overset{6}{\geq}
    	\ln_r
    	\left[
    	p_q
    	\left(
    	XG
    	\right)^{-1}
    	K^{-1}
    	\right]
    	.
    \end{align}
    In the fourth line we use the property of pseudo-substraction. We now use that the r-deformed logarithm is increasing and get:
    \begin{align}
        p_q
    	\left(
    	X|G
    	\right)^{-1}
        &\geq
        p_q
    	\left(
    	XG
    	\right)^{-1}
    	K^{-1}
    	,
    	\\
        p_q
    	\left(
    	XG
    	\right)
    	K
    	&\geq
    	p_q
    	\left(
    	X|G
    	\right)
    	.
    \end{align}
    This resulting inequality has been proven for $q\geq0$ \cite{review_RCE}. If we consider the remaining case $q < 0$, the same arguments lead instead to the inequality 
    $p_q
	(
	X|G
	)
	\geq
	p_q
	(
	XG
	)
	K$. We can prove this latter inequality using similar arguments as for the positive case, and we do it below for completeness.
    \begin{align}
    &p_q
    	\left(
    	X|G
    	\right)
    \nonumber \\
    &\overset{1}{=}
    \left(
	\sum_g
	\left(
	\sum_x
	p(x,g)^q
	\right)^\frac{1}{q}	
	\right)^\frac{q}{(q-1)}
	\hspace{-0.6cm}
	,
	\\
	&\overset{2}{=}
    \left(
    K
	\sum_g
	\frac{1}{K}
	\left(
	\sum_x
	p(x,g)^q
	\right)^\frac{1}{q}	
	\right)^\frac{q}{(q-1)}
	\hspace{-0.6cm}
	,
	\\
	&\overset{3}{\geq}
    \left(
    K
    \left(
	\sum_g
    \frac{1}{K}
	\sum_x
	p(x,g)^q
	\right)^\frac{1}{q}	
	\right)^\frac{q}{(q-1)}
	\hspace{-0.6cm}
	,
	\\
	&\overset{4}{=}
	K^\frac{q}{q-1}
	\left(\frac{1}{K}\right)^\frac{1}{q-1}
	\hspace{-0.1cm}
    \left(
    \left(
	\sum_{g,x}
	p(x,g)^q
	\right)^\frac{1}{q}	
	\right)^\frac{q}{(q-1)}
	\hspace{-0.7cm}
	,
	\\
	&\overset{5}{=}
	K\,
    p_q(XG)
	.
\end{align}
In the first line we invoke the definition of the Arimoto-R\'enyi quantity. In the second line we introduce the factor $1/K$ so to use Jensen's inequality. In the third line we use Jensen's inequality, taking into account that the function $x^\frac{1}{q}$ is convex for $q<0$, and that $\frac{q}{q-1}>0$ for $q<0$, so it does not change the order of the inequality after using Jensen's inequality. In the fourth line we take out the constants. In the fifth line we simplify the expression and identify the R\'enyi quantity \eqref{eq:rp}. This completes the proof.
\end{proof}

\newpage
\twocolumngrid
\bibliographystyle{apsrev4-1}
\bibliography{bibliography.bib}

\end{document}